\documentclass[aps,onecolumn,floatfix,12pt]{revtex4}

\usepackage{epsfig}
\usepackage{amssymb}
\usepackage{amsmath}
\usepackage{amsfonts}
\usepackage{epstopdf}
\usepackage{verbatim}
\usepackage{color}
\usepackage[normalem]{ulem}
\usepackage[english]{babel}
\allowdisplaybreaks

\begin{document}

\title{Fluid kinetic energy asymptotic expansion for two variable radii moving spherical bubbles at small separation distance}

\author{ S.\,V.~Sanduleanu$^{a,b}$\thanks {e-mail:shtefan.sanduleanu@gmail.com}}

\affiliation{$^a$Ishlinsky Institute for Problems in Mechanics RAS, Moscow 119526, Russia}

\affiliation{$^b$Moscow Institute of Physics and Technology, Moscow 141700, Russia}

\begin{abstract}
Two spherical bubbles with changing radii are considered to be moving in ideal fluid along their center-line. The exact expression for the fluid kinetic energy is obtained. The Stokes stream function is expanded in Gegenbauer polynomials in bispherical coordinates. This expansion is used to obtain the exact series for the fluid kinetic energy quadratic form coefficients. The new series are confirmed to be correct by comparison with the known ones. The main advantage of the new kinetic energy form is the possibility to obtain asymptotic expansions at small separation distance between the bubbles. These expansions are obtained and their convergence is analyzed. The results of this work can be used to describe the bubbles approach before the contact and their coalescence in acoustic field.
\end{abstract}

\maketitle
\section{Introduction}
The problem of interaction of two spherical gas bubbles in fluid in an acoustic  field is the object of study of numerous theoretical and practical works, starting with Bjerknes's works in the 19th century \cite{bjerknes1906fields}. This problem is being considered in many contemporary works \cite{zilonova2019dynamics,doinikov2015theoretical,cleve2018surface,jiao2015experimental}. Bjerknes determined that the interaction force between two pulsating spheres, the distance between which is rather large, is inversely proportional to the square of the distance between the spheres. This dependence was proved experimentally in \cite{kazantsev1960motion,crum1975bjerknes}. However, both in theoretical \cite{doinikov2015theoretical,petrov2011forced} and experimental works \cite{jiao2013experimental,jiao2015experimental,jiao2015influence,garbin2007changes} demonstrated the inapplicability of this dependence near the contact. It should be found from the solution of two pulsating spheres problem in the exact formulation.

The generalized Lagrange coordinates proved to be convenient to study the problem of interaction of two gas bubbles in an acoustic field. The main summand of the Lagrange function is the kinetic energy.
The kinetic energy may be calculated in terms of the following parameters: sphere radii, their change rate in time, the distance between the spheres' centers and the centers' velocities.

To construct the exact expression for the kinetic energy, there exist two methods, considered most effective. The first is the reflection method, which was developed by Hicks in the classic work \cite{hicks1880}. He built the exact solution for the motion of two solid spheres along their centerlines. The kinetic energy is the quadratic form of the centers' velocities. For the coefficients, Hicks obtained series that converge absolutely for any values of the geometric parameters of the problem and are very useful for calculating the coefficients with any required accuracy. Using these series, Voinov \cite{Voinov1969PMM} found the three term asymptotic expansion of coefficients at small separation distance.

The second method for solving this problem was suggested by Neumann in \cite{neumann1883hydrodynamische}. He expressed the velocity potential in bispherical coordinates. He obtained the same series as Hicks for the kinetic energy coefficients. Neumann also transformed these series to another form. Rasziilier \emph{et al.} \cite{raszillier1990optimal} used the second form of Neumann series to construct the asymptotic expansion at small separation distance. Bentwich and Miloh \cite{bentwich1978exact} obtained the second form of Neumann series, solving the problem in bispherical coordinates for the Stokes stream function by Jeffrey's method \cite{jeffery1912form}.

For the first time, the exact solution for the problem with variable radii was obtained by Voinov  in \cite{Voinov1970} using Hicks's reflection method \cite{hicks1880,hicks1879pt1,hicks1879pt2}. Although Selby \cite{selby1890} found earlier the approximated kinetic energy using several reflections. Besides three Hicks's coefficients, the quadratic form contains seven additional coefficients, which Voinov presented in the form of series, similar to Hicks's series \cite{Voinov1970,Voinov1969vestnik,VoinovPetrov1976}. In some cases, Voinov suggested a method for determining the first coefficients of asymptotic series \cite{Voinov1970,Voinov1969PMM}. Developing Voinov's ideas, in \cite{sanduleanu2018trinomial} three-terms asymptotic expansions  at small separation distance  were found for all of the ten coefficients.

There also exists a series of works, in which the exact solution is constructed by the inverse powers of the distance between sphere centers  $r$. Such solutions have a more complex form and their applicability at small separation distances is questionable. In \cite{kuznetsov1972interaction} and \cite{doinikov2001translational} the kinetic energy is found up to  ${{r}^{-3}}$. In \cite{harkin2001coupled} the velocity potential expansion up to ${{r}^{-4}}$  apparently contains errors which were noticed in \cite{doinikov2015theoretical}. In \cite{aganin2009refined} the solution is built with accuracy  ${{r}^{-5}}$. In \cite{doinikov2015theoretical}  a solution is presented for which can be shown that the accuracy does not exceed  ${{r}^{-6}}$.

In recent works \cite{maksimov2016coupled,maksimov2018scattering},  based on \cite{morioka1974theory}, the interaction of two bubbles of varying radii with fixed centers was considered. The coefficients of the kinetic energy differ from the exact ones. This difference is explained by the rough assumption that the velocity potential is constant on the sphere surface. This inaccuracy does not influence the main asymptotics of the secondary Bjerknes force for  large distances between the spheres centers.

Thus, currently there is no precise asymptotic expansion for the fluid kinetic energy for bubbles near the contact. Such an expansion is necessary for describing the process of bubbles approach and the analysis of possibility of their coalescence. In this work two spherical bubbles with changing radii, moving along their centerline, are considered and the asymptotic expansion by a small separation distance for the fluid kinetic energy is obtained. The convergence of the asymptotic expansion is studied.

\section{Kinetic energy }
\subsection{Problem formulation}
\begin{figure}[ht]
\includegraphics{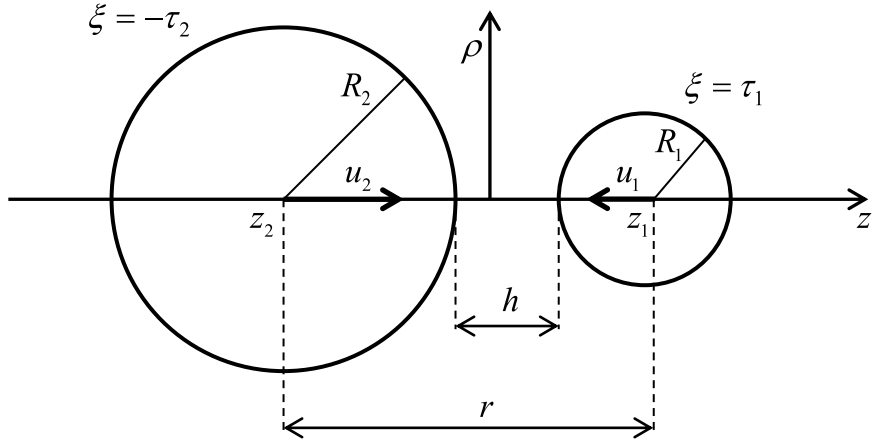}
\caption{Problem formulation.}
\label{fig1}
\vspace{0.0 in}
\end{figure}

We consider the problem of finding the kinetic energy of potential axially symmetric flow in infinite incompressible fluid of density ${{\rho }_{l}}$. The fluid flow is caused by two spheres of radii ${{R}_{1}},{{R}_{2}}$ changing with velocities ${{\dot{R}}_{1}},{{\dot{R}}_{2}}$.
The centers of spheres are situated on axis  $z$ at ${{z}_{1}},{{z}_{2}},({{z}_{1}}>{{z}_{2}})$. They move with velocities ${{u}_{1}}=-{{\dot{z}}_{1}}$, ${{u}_{2}}={{\dot{z}}_{2}}$ directed towards each other (Fig.\ref{fig1}). The distance between the spheres centers is  $r={{z}_{1}}-{{z}_{2}}$, the distance between the spheres surfaces is  $h=r-{{R}_{1}}-{{R}_{2}}$. The goal of this work is to find the analytic dependence of the kinetic energy  $T({{R}_{1}},{{R}_{2}},r,{{u}_{1}},{{u}_{2}},{{\dot{R}}_{1}},{{\dot{R}}_{2}})$. With such choice of arguments, the kinetic energy is symmetric with respect to the substricts $1$ and $2$ permutation.

The fluid velocity components  ${{v}_{\rho }},{{v}_{\theta }},{{v}_{z}}$, in the cylindrical coordinates system ($\rho ,\theta ,z$)  are expressed through the Stokes stream function $\psi $

\begin{equation}\label{velStokescyl}
\begin{aligned}
&{{v}_{\rho }}=\frac{1}{\rho }\frac{\partial \psi }{\partial z},\\
&{{v}_{\theta }}=0,\\
&{{v}_{z}}=-\frac{1}{\rho }\frac{\partial \psi }{\partial \rho }.\\
\end{aligned}
\end{equation}
The equation for the stream function $\psi $ has the following form \cite{lamb1993hydrodynamics}
\begin{equation}\label{StreamFunctionEqCyl}
\frac{\partial }{\partial \rho }\left( \frac{1}{\rho }\frac{\partial \psi }{\partial \rho } \right)+\frac{\partial }{\partial z}\left( \frac{1}{\rho }\frac{\partial \psi }{\partial z} \right)=0
\end{equation}

\subsection{Bispherical coordinates }
It is convenient to consider the bispherical coordinates ($\xi ,\zeta ,\theta $)
\begin{equation}\label{bispherecoord}
\begin{aligned}
&\rho =c\frac{\sin \zeta }{\cosh \xi -\cos \zeta } \\
&z=c\frac{\sinh \xi }{\cosh \xi -\cos \zeta } \\
&(x=\rho \cos \theta ,y=\rho \sin \theta ) \\
\end{aligned}
\end{equation}
Then the surface of the first sphere of radius ${{R}_{1}}$  is given by the following equation
\begin{equation}\label{sphere1}
\xi ={{\tau }_{1}}=const,\,\,\,\,\,\,\zeta \in [0,\pi ],\theta \in [0,2\pi ],
\end{equation}
 the surface of the second bubble of radius ${{R}_{2}}$  is given by
\begin{equation}\label{sphere2}
\xi =-{{\tau }_{2}}=const,\,\,\,\,\,\zeta \in [0,\pi ],\theta \in [0,2\pi ],
\end{equation}
and
\begin{equation}\label{eqfort1t2c}
\begin{aligned}
&{{R}_{1}}\sinh {{\tau }_{1}}=c,\\
&{{R}_{2}}\sinh {{\tau }_{2}}=c,\\
& r={{R}_{1}}\cosh {{\tau }_{1}}+{{R}_{2}}\cosh {{\tau }_{2}}.  \\
\end{aligned}
\end{equation}
Thus, we determine the spheres surfaces with the help of parameters  ${{\tau }_{1}},{{\tau }_{2}}$ and  $c$, which can be expressed through  ${{R}_{1}},{{R}_{2}}$ and small separation distance $h$
\begin{equation}\label{t1plust2}
\begin{aligned}
&{{\tau}_{1}}=\frac{{{R}_{2}}}{{{R}_{1}}+{{R}_{2}}}\sqrt{\frac{2h}{p}}+O\left( {{h}^{3/2}} \right),\\
&{{\tau }_{2}}=\frac{{{R}_{1}}}{{{R}_{1}}+{{R}_{2}}}\sqrt{\frac{2h}{p}}+O\left( {{h}^{3/2}} \right),\\
&c=\sqrt{2hp}+O\left( {{h}^{3/2}} \right),\\
&p=\frac{{{R}_{1}}{{R}_{2}}}{{{R}_{1}}+{{R}_{2}}}.  \\
\end{aligned}
\end{equation}

\subsection{Stream function }
To find the stream function, we write down the potential flow equation (\ref{StreamFunctionEqCyl}) in bispheric coordinates \cite{jeffery1912form}

\begin{equation}\label{StreamFunctionEqBisher}
\frac{\partial }{\partial \xi }\left( \frac{1}{\rho }\frac{\partial \psi }{\partial \xi } \right)+\frac{\partial }{\partial \zeta }\left( \frac{1}{\rho }\frac{\partial \psi }{\partial \zeta } \right)=0.
\end{equation}

The stream function can be presented as follows \cite{jeffery1912form}
\begin{equation}\label{StreamFunctionSol}
\begin{aligned}
  & \psi ={{\left( \cosh \xi -\cos \zeta  \right)}^{-1/2}}\sum\limits_{n=0}^{\infty }{{{U}_{n}}(\xi )C_{n}^{-1/2}(\mu )}, \\
 &  {{U}_{n}}(\xi )=\frac{{{\alpha }_{n}}\sinh (n-1/2)(\xi +{{\tau }_{2}})+{{\beta }_{n}}\sinh (n-1/2)({{\tau }_{1}}-\xi )}{\sinh (n-1/2)({{\tau }_{1}}+{{\tau }_{2}})},  \\
\end{aligned}
\end{equation}
where  $C_{n}^{-1/2}(\mu )$ are the Gegenbauer polynomials, $\mu =\cos \zeta $  (it is sometimes convenient to use this substitution). The Gegenbauer polynomials can be obtained from the following recurrent relation \cite{whittaker1996course}

\begin{equation}\label{Gegerecurr}
nC_{n}^{-1/2}(\mu )=2\mu (n-3/2)C_{n-1}^{-1/2}(\mu )-(n-3)C_{n-2}^{-1/2}(\mu ),~~~~ C_{0}^{-1/2}(\mu )=1,C_{1}^{-1/2}(\mu )=-\mu .
\end{equation}

The coefficients ${{\alpha }_{n}},{{\beta }_{n}}$ can be obtained from boundary conditions on spheres surfaces for  $\xi ={{\tau }_{1}}$ and  $\xi =-{{\tau }_{2}}$. They may be written as follows

\begin{equation}\label{GranEq}
\begin{aligned}
  & ({{\mathbf{v}}_{1}},\mathbf{n})=\frac{1}{\rho }\frac{\partial \psi }{\partial \zeta }\frac{\cosh {{\tau }_{1}}-\cos \zeta }{c}=-{{u}_{1}}\left( -\frac{\cosh {{\tau }_{1}}\cos \zeta -1}{\cosh {{\tau }_{1}}-\cos \zeta } \right)-{{\dot{R}}_{1}}, \\
 &  ({{\mathbf{v}}_{2}},\mathbf{n})=\frac{1}{\rho }\frac{\partial \psi }{\partial \zeta }\frac{\cosh {{\tau }_{2}}-\cos \zeta }{c}={{u}_{2}}\left( -\frac{\cosh {{\tau }_{2}}\cos \zeta -1}{\cosh {{\tau }_{2}}-\cos \zeta } \right)+{{\dot{R}}_{2}}.  \\
\end{aligned}
\end{equation}

Integrating both boundary conditions by  $\zeta $ and choosing the integration constants so that on the symmetry axis the velocity is parallel to this axis, the following relations are obtained

\begin{equation}\label{GranValueStreamFunc}
\begin{aligned}
  & {{\left. \psi  \right|}_{\xi ={{\tau }_{1}}}}={{u}_{1}}{{c}^{2}}\frac{1}{2}\frac{1-{{\mu }^{2}}}{{{\left( \cosh {{\tau }_{1}}-\mu  \right)}^{2}}}+{{{\dot{R}}}_{1}}{{c}^{2}}\frac{1}{\cosh {{\tau }_{1}}-\mu }+\left( \frac{{{{\dot{R}}}_{2}}{{c}^{2}}}{{{\sinh }^{2}}{{\tau }_{2}}}-\frac{{{{\dot{R}}}_{1}}{{c}^{2}}\cosh {{\tau }_{1}}}{{{\sinh }^{2}}{{\tau }_{1}}} \right), \\
 & {{\left. \psi  \right|}_{\xi =-{{\tau }_{2}}}}=-{{u}_{2}}{{c}^{2}}\frac{1}{2}\frac{1-{{\mu }^{2}}}{{{\left( \cosh {{\tau }_{2}}-\mu  \right)}^{2}}}-{{{\dot{R}}}_{2}}{{c}^{2}}\frac{1}{\cosh {{\tau }_{2}}-\mu }-\left( \frac{{{{\dot{R}}}_{1}}{{c}^{2}}}{{{\sinh }^{2}}{{\tau }_{1}}}-\frac{{{{\dot{R}}}_{2}}{{c}^{2}}\cosh {{\tau }_{2}}}{{{\sinh }^{2}}{{\tau }_{2}}} \right). \\
\end{aligned}
\end{equation}

Substituting the general expression for current function (\ref{StreamFunctionSol}) in the boundary conditions, the following system is obtained

\begin{equation}\label{StreamFunctionCoef}
\begin{aligned}
 \sum\limits_{n=0}^{\infty }{{{\alpha }_{n}}C_{n}^{-1/2}(\mu )}=&{{u}_{1}}{{c}^{2}}\frac{1}{2}\frac{1-{{\mu }^{2}}}{{{\left( \cosh {{\tau }_{1}}-\mu  \right)}^{3/2}}}+{{\dot{R}}_{1}}{{c}^{2}}\frac{1}{{{\left( \cosh {{\tau }_{1}}-\mu  \right)}^{1/2}}}\\
&+\left( \frac{{{{\dot{R}}}_{2}}{{c}^{2}}}{{{\sinh }^{2}}{{\tau }_{2}}}-\frac{{{{\dot{R}}}_{1}}{{c}^{2}}\cosh {{\tau }_{1}}}{{{\sinh }^{2}}{{\tau }_{1}}} \right){{\left( \cosh {{\tau }_{1}}-\mu  \right)}^{1/2}},\\
 \sum\limits_{n=0}^{\infty }{{{\beta }_{n}}C_{n}^{-1/2}(\mu )}=&-{{u}_{2}}{{c}^{2}}\frac{1}{2}\frac{1-{{\mu }^{2}}}{{{\left( \cosh {{\tau }_{2}}-\mu  \right)}^{3/2}}}-{{\dot{R}}_{2}}{{c}^{2}}\frac{1}{{{\left( \cosh {{\tau }_{2}}-\mu  \right)}^{1/2}}}\\
&-\left( \frac{{{{\dot{R}}}_{1}}{{c}^{2}}}{{{\sinh }^{2}}{{\tau }_{1}}}-\frac{{{{\dot{R}}}_{2}}{{c}^{2}}\cosh {{\tau }_{2}}}{{{\sinh }^{2}}{{\tau }_{2}}} \right){{\left( \cosh {{\tau }_{2}}-\mu  \right)}^{1/2}} . \\
\end{aligned}
\end{equation}

To find the coefficients  ${{\alpha }_{n}},{{\beta }_{n}}$, we expand the right sides of equations in Gegenbauer polynomials, using (Appendix \ref{AppendixGranCond})

\begin{equation}\label{GranCond1Gege}
{{\left( \cosh {{\tau }_{1}}-\mu  \right)}^{1/2}}=\sum\limits_{n=0}^{\infty }{C_{n}^{-1/2}(\mu )\frac{\sqrt{2}}{2}{{e}^{-(n-1/2){{\tau }_{1}}}}},
\end{equation}

\begin{equation}\label{GranCond2Gege}
\frac{1}{{{\left( \cosh {{\tau }_{1}}-\mu  \right)}^{1/2}}}=\sum\limits_{n=0}^{\infty }{C_{n}^{-1/2}(\mu )\frac{-(n-1/2)\sqrt{2}}{\sinh {{\tau }_{1}}}{{e}^{-(n-1/2){{\tau }_{1}}}}},
\end{equation}

\begin{equation}\label{GranCond3Gege}
\frac{1}{2}\frac{1-{{\mu }^{2}}}{{{\left( \cosh {{\tau }_{1}}-\mu  \right)}^{3/2}}}=\sum\limits_{n=0}^{\infty }{C_{n}^{-1/2}(\mu )\sqrt{2}n(n-1){{e}^{-(n-1/2){{\tau }_{1}}}}}.
\end{equation}

After substituting (\ref{GranCond1Gege}-\ref{GranCond3Gege}) in (\ref{StreamFunctionCoef}) we get the final expression for the stream function

\begin{equation}\label{StreamFunctionSolFinal}
\begin{aligned}
   \psi =&{{\left( \cosh \xi -\cos \zeta  \right)}^{-1/2}}\sum\limits_{n=0}^{\infty }{{{U}_{n}}(\xi )C_{n}^{-1/2}(\mu )},\\
  {{U}_{n}}(\xi )=&\frac{{{\alpha }_{n}}\sinh (n-1/2)(\xi +{{\tau }_{2}})+{{\beta }_{n}}\sinh (n-1/2)({{\tau }_{1}}-\xi )}{\sinh (n-1/2)({{\tau }_{1}}+{{\tau }_{2}})},  \\
  {{\alpha }_{n}}=&\left( {{u}_{1}}2n(n-1)+{{{\dot{R}}}_{1}}\frac{-(2n-1)}{\sinh {{\tau }_{1}}}+\left( \frac{{{{\dot{R}}}_{2}}}{{{\sinh }^{2}}{{\tau }_{2}}}-\frac{{{{\dot{R}}}_{1}}\cosh {{\tau }_{1}}}{{{\sinh }^{2}}{{\tau }_{1}}} \right) \right)\frac{{{c}^{2}}\sqrt{2}}{2}{{e}^{-(n-1/2){{\tau }_{1}}}},\\
  {{\beta }_{n}}=&\left( -{{u}_{2}}2n(n-1)-{{{\dot{R}}}_{2}}\frac{-(2n-1)}{\sinh {{\tau }_{2}}}-\left( \frac{{{{\dot{R}}}_{1}}}{{{\sinh }^{2}}{{\tau }_{1}}}-\frac{{{{\dot{R}}}_{2}}\cosh {{\tau }_{2}}}{{{\sinh }^{2}}{{\tau }_{2}}} \right) \right)\frac{{{c}^{2}}\sqrt{2}}{2}{{e}^{-(n-1/2){{\tau }_{2}}}}.\\
\end{aligned}
\end{equation}

\subsection{Kinetic energy}
The kinetic energy is expressed through the integral of  ${{v}^{2}}$ on the domain outside two spheres

\begin{equation}\label{KinInt1}
T=\frac{{{\rho }_{l}}}{2}\iiint{{{v}^{2}}dV}.
\end{equation}
It can be rewritten as follows

\begin{equation}\label{vsqdV}
\begin{aligned}
 {{v}^{2}}dV & =\left( {{\left( \frac{1}{\rho }\frac{\cosh \xi -\cos\zeta }{c}\frac{\partial \psi }{\partial \xi } \right)}^{2}}+{{\left( \frac{1}{\rho }\frac{\cosh \xi -\cos\zeta }{c}\frac{\partial \psi }{\partial \zeta } \right)}^{2}} \right){{\left( \frac{c}{\cosh \xi -\cos\zeta } \right)}^{2}}\rho \,d\theta d\xi d\zeta \\
& =\frac{1}{\rho }\left( {{\left( \frac{\partial \psi }{\partial \xi } \right)}^{2}}+{{\left( \frac{\partial \psi }{\partial \zeta } \right)}^{2}} \right)d\theta d\xi d\zeta.
\end{aligned}
\end{equation}
Integrating by $\theta $, we obtain that

\begin{equation}\label{KinInt1simple}
T=\frac{{{\rho }_{l}}}{2}\iint{2\pi \left( \psi {{'}_{\xi }}\frac{1}{\rho }\psi {{'}_{\xi }}+\psi {{'}_{\zeta }}\frac{1}{\rho }\psi {{'}_{\zeta }} \right)}\,d\xi d\zeta.
\end{equation}
Taking into account the potentiality of flow (\ref{StreamFunctionEqBisher}), we get the following

\begin{equation}\label{dpsirodpsi}
\psi {{'}_{\xi }}\frac{1}{\rho }\psi {{'}_{\xi }}+\psi {{'}_{\zeta }}\frac{1}{\rho }\psi {{'}_{\zeta }}=\frac{\partial }{\partial \xi }\left( \psi \frac{1}{\rho }\psi {{'}_{\xi }} \right)+\frac{\partial }{\partial \zeta }\left( \psi \frac{1}{\rho }\psi {{'}_{\zeta }} \right).
\end{equation}
Using this equality and Green's formula, the kinetic energy can be found as follows \cite{bentwich1978exact}

\begin{equation}\label{KineticIntegrContour}
\frac{T}{\pi {{\rho }_{l}}}=\oint{\psi \frac{1}{\rho }\psi {{'}_{\zeta }}d\xi -\psi \frac{1}{\rho }\psi {{'}_{\xi }}d\zeta }=\int\limits_{\xi =-{{\tau }_{2}}}^{{{\tau }_{1}}}{\left. \psi \frac{1}{\rho }\psi {{'}_{\zeta }} \right|_{\zeta =0}^{\zeta =\pi }d\xi }+\int\limits_{\zeta =0}^{\pi }{\left. \psi \frac{1}{\rho }\psi {{'}_{\xi }} \right|_{\xi =-{{\tau }_{2}}}^{\xi ={{\tau }_{1}}}d\zeta }.
\end{equation}
Note that in the case of constant radii spheres, the stream function equals zero on the symmetry axis (see \cite{bentwich1978exact}). Thus, in this case the first integral equals zero.

In the case considered in this paper, the spheres radii are variable, thus the first integral should be preserved. Indeed, taking into account that $C_{n}^{-1/2}(\cos \zeta )=0$ as  $\zeta =\{0,\pi \}$  for  $n\ge 2$,     $C_{0}^{-1/2}(\cos \zeta )=1$,
$C_{1}^{-1/2}(\cos \zeta )=-\cos \zeta $, we obtain that

\begin{equation}\label{psionaxis}
\begin{aligned}
  & {{\left. \psi  \right|}_{\zeta =\pi }}=\frac{1}{\sqrt{\cosh \xi +1}}\left( {{U}_{0}}\left( \xi  \right)+{{U}_{1}}\left( \xi  \right) \right)={{c}^{2}}\left( \frac{{{{\dot{R}}}_{2}}}{{{\sinh }^{2}}{{\tau }_{2}}}-\frac{{{{\dot{R}}}_{1}}}{{{\sinh }^{2}}{{\tau }_{1}}} \right) \\
 & {{\left. \psi  \right|}_{\zeta =0}}=\frac{1}{\sqrt{\cosh \xi -1}}\left( {{U}_{0}}\left( \xi  \right)-{{U}_{1}}\left( \xi  \right) \right)={{c}^{2}}\left( \frac{{{{\dot{R}}}_{2}}}{{{\sinh }^{2}}{{\tau }_{2}}}+\frac{{{{\dot{R}}}_{1}}}{{{\sinh }^{2}}{{\tau }_{1}}} \right)\frac{\sinh \xi /2}{\left| \sinh \xi /2 \right|} \\
\end{aligned}
\end{equation}
Taking into consideration that $\frac{dC_{n}^{-1/2}(\mu )}{d\mu }=-1$  as  $\mu =1$ and $\frac{dC_{n}^{-1/2}(\mu )}{d\mu }={{(-1)}^{n}}$    as $\mu =-1$  for $n\ge 1$, finding the first integral of (\ref{KineticIntegrContour}) is easy.
The second integral may be found by substituting  $\psi $ into the boundary  conditions(\ref{GranValueStreamFunc}),  $\psi {{'}_{\xi }}$ is found using (\ref{StreamFunctionSol}) and calculating the necessary integrals (see Appendix \ref{AppendixI1I6}).

After some transformations (see Appendix \ref{AppendixA1E}), we simplify the formulas of kinetic energy

\begin{equation}\label{KinEn}
\begin{aligned}
  & T=2\pi {{\rho }_{l}}\big( {{A}_{1}}u_{1}^{2}+2B{{u}_{1}}{{u}_{2}}+{{A}_{2}}u_{2}^{2}+{{D}_{1}}{{{\dot{R}}}_{1}}^{2}+2E{{{\dot{R}}}_{1}}{{{\dot{R}}}_{2}}+{{D}_{2}}{{{\dot{R}}}_{2}}^{2}\\
&~~~~~+{{C}_{11}}{{u}_{1}}{{{\dot{R}}}_{1}}+{{C}_{12}}{{u}_{1}}{{{\dot{R}}}_{2}}+{{C}_{21}}{{u}_{2}}{{{\dot{R}}}_{1}}+{{C}_{22}}{{u}_{2}}{{{\dot{R}}}_{2}} \big), \\
 &{{A}_{1}}=\frac{R_{1}^{3}}{6}+{{c}^{3}}\sum\limits_{n=2}^{\infty }{\frac{{{e}^{-\left( 2n-1 \right){{\tau }_{1}}}}}{{{e}^{\left( 2n-1 \right)\left( {{\tau }_{1}}+{{\tau }_{2}} \right)}}-1}}\frac{{{(2n-1)}^{2}}-1}{2}, \\
  &B={{c}^{3}}\sum\limits_{n=2}^{\infty }{\frac{1}{{{e}^{\left( 2n-1 \right)\left( {{\tau }_{1}}+{{\tau }_{2}} \right)}}-1}\frac{{{(2n-1)}^{2}}-1}{2}}, \\
  &{{C}_{11}}=2{{c}^{3}}\sum\limits_{n=2}^{\infty }{\frac{{{e}^{-(2n-1){{\tau }_{1}}}}}{{{e}^{\left( 2n-1 \right)\left( {{\tau }_{1}}+{{\tau }_{2}} \right)}}-1}{{S}_{n}}({{\tau }_{1}}),}\\
& {{C}_{12}}=2{{c}^{3}}\sum\limits_{n=2}^{\infty }{\frac{1}{{{e}^{\left( 2n-1 \right)\left( {{\tau }_{1}}+{{\tau }_{2}} \right)}}-1}}{{S}_{n}}({{\tau }_{2}}), \\
&  {{D}_{1}}=R_{1}^{3}+{{c}^{3}}\sum\limits_{n=2}^{\infty }{\frac{{{e}^{-(2n-1){{\tau }_{1}}}}}{{{e}^{\left( 2n-1 \right)\left( {{\tau }_{1}}+{{\tau }_{2}} \right)}}-1}}\frac{2\,S_{n}^{2}({{\tau }_{1}})}{{{(2n-1)}^{2}}-1},\\
& E=\frac{{{({{R}_{1}}{{R}_{2}})}^{2}}}{r}+{{c}^{3}}\sum\limits_{n=2}^{\infty }{\frac{{{e}^{-(2n-1)({{\tau }_{1}}+{{\tau }_{2}})}}}{{{e}^{\left( 2n-1 \right)\left( {{\tau }_{1}}+{{\tau }_{2}} \right)}}-1}}\frac{2\,{{S}_{n}}({{\tau }_{1}}){{S}_{n}}({{\tau }_{2}})}{{{(2n-1)}^{2}}-1}, \\
\end{aligned}
\end{equation}
where  ${{S}_{n}}(x)=\frac{{{e}^{(2n-1)x}}-(2n-1)\sinh x-\cosh x}{{{\sinh }^{2}}x}$, and the coefficients  ${{A}_{2}},{{C}_{21}},{{C}_{22}}$  and ${{D}_{2}}$   are obtained by permuting subscripts 1 and 2 in formulas for  ${{A}_{1}},{{C}_{12}},{{C}_{11}},{{D}_{1}}$. Moreover, the series (\ref{KinEn}) may be expressed through the initial parameters  ${{R}_{1}},{{R}_{2}},r$, using the following substitutions

\begin{equation}\label{expt1t2c}
\begin{aligned}
   {{e}^{-{{\tau }_{1}}}}=&\frac{{{r}^{2}}+R_{1}^{2}-R_{2}^{2}}{2r{{R}_{1}}}-\sqrt{{{\left( \frac{{{r}^{2}}+R_{1}^{2}-R_{2}^{2}}{2r{{R}_{1}}} \right)}^{2}}-1}, \\
 {{e}^{-{{\tau}_{2}}}}=&\frac{{{r}^{2}}+R_{2}^{2}-R_{1}^{2}}{2r{{R}_{2}}}-\sqrt{{{\left( \frac{{{r}^{2}}+R_{2}^{2}-R_{1}^{2}}{2r{{R}_{2}}} \right)}^{2}}-1}, \\
  {{e}^{-({{\tau }_{1}}+{{\tau}_{2}})}}=&\frac{{{r}^{2}}-R_{1}^{2}-R_{2}^{2}}{2{{R}_{1}}{{R}_{2}}}-\sqrt{{{\left( \frac{{{r}^{2}}-R_{1}^{2}-R_{2}^{2}}{2{{R}_{1}}{{R}_{2}}} \right)}^{2}}-1} ,\\
  c=&\frac{\sqrt{{{\left( {{r}^{2}}-R_{1}^{2}-R_{2}^{2} \right)}^{2}}-4R_{1}^{2}R_{2}^{2}}}{2r}. \\
\end{aligned}
\end{equation}
\subsection{Hicks and Voinov series }
In case of solid spheres, the exact expression of kinetic energy was first found by Hicks in \cite{hicks1880}, using the reflection method. It is described in Lamb's monography \cite{lamb1993hydrodynamics}. O.V. Voinov developed Hicks's method for the case of varying radii. The kinetic energy coefficients, found by Hicks, have the following form

\begin{equation}\label{KinEnHicks}
{{A}_{1}}=\frac{R_{1}^{3}}{6}+\frac{1}{2}\sum\limits_{j=1}^{\infty }{{{\left( \frac{{{R}_{1}}}{A_{j}^{1}} \right)}^{3}}}, ~~~~~~~ B=\frac{1}{2}\sum\limits_{j=1}^{\infty }{{{\left( \frac{{{R}_{2}}}{B_{j}^{1}} \right)}^{3}}}.
\end{equation}

Voinov obtained the rest of the coefficients (detailed derivation can be found in \cite{petrov2011forced}))

\begin{equation}\label{KinEnVoinov}
\begin{aligned}
& {{C}_{11}}=\sum\limits_{j=1}^{\infty }{\frac{R_{1}^{3}}{{{\left( A_{j}^{1} \right)}^{2}}B_{j}^{2}}},~~~~~~~{{C}_{12}}=\sum\limits_{j=1}^{\infty }{\frac{R_{1}^{3}}{{{\left( B_{j}^{2} \right)}^{2}}A_{j-1}^{1}}},\\
& {{D}_{1}}=R_{1}^{3}+\sum\limits_{j=1}^{\infty }{\frac{R_{1}^{3}}{A_{j}^{1}}\left[ 1+({{(B_{j}^{2})}^{2}}-1)\ln \left( 1-\frac{1}{{{(B_{j}^{2})}^{2}}} \right) \right]},\\
&E=\frac{{{({{R}_{1}}{{R}_{2}})}^{2}}}{r}+\frac{{{({{R}_{1}}{{R}_{2}})}^{2}}}{{{R}_{2}}}\sum\limits_{j=1}^{\infty }{\left( \frac{1}{B_{j+1}^{2}}-B_{j}^{1}\ln \left( 1+\frac{1}{B_{j+1}^{2}B_{j}^{1}} \right) \right)},\\
\end{aligned}
\end{equation}
where  $A_{j}^{i},B_{j}^{i}$ can be found from the recurrent formulas
\begin{equation}\label{recurrVoinov}
B_{j}^{i}=\frac{r}{{{R}_{i}}}A_{j-1}^{k}-\frac{{{R}_{k}}}{{{R}_{i}}}B_{j-1}^{i},A_{j}^{i}=\frac{r}{{{R}_{i}}}B_{j}^{k}-\frac{{{R}_{k}}}{{{R}_{i}}}A_{j-1}^{i},
~~~~~~~i,k=1,2,~~~~~~\,i\ne k,
\end{equation}
with initial conditions $A_{0}^{i}=1,\,\,B_{0}^{i}=0$.

These series are expressed through parameters  ${{R}_{1}},{{R}_{2}},r$. For $r>{{R}_{1}}+{{R}_{2}}$  the series converge as geometric progression. In case of contact  $r={{R}_{1}}+{{R}_{2}}$ they converge as power series ($1/{{n}^{3}}$). However, the derivates of these series, which are necessary for calculating the forces, diverge when approaching contact. But series  (\ref{KinEn}) allow one to obtain the expansion in the small parameter $h$. The derivatives of these expansions contain a logarithmic singularity. The asymptotic expansion of the interaction force, obtained from series (\ref{KinEn}), allows the analytical study  of bubbles approach up to the contact point.

\subsection{Comparison of kinetic energy expression}
Although the Hicks (\ref{KinEnHicks}) and Voinov (\ref{KinEnVoinov}) series seem to differ from series (\ref{KinEn}), they are identical. This fact is shown in Appendix \ref{AppendixEqualCoef}, which verifies both results.
	In \cite{doinikov2015theoretical} the kinetic energy is found as infinite sums by the inverse powers of $r$. The comparison with the exact solution shows that summands up to ${{r}^{-6}}$ coincide, but the following ones don't (see Appendix \ref{AppendixKinEnSeries}).

\section{Asymptotic expansion}
\subsection{Asymptotic expansion at small separation distance}
To obtain the asymptotic expansion of fluid kinetic energy at small separation distance, we use the method described in Raszillier \emph{et al.} \cite{raszillier1989short} (for spheres of equal radii), Raszillier \emph{et al.} \cite{raszillier1990optimal} (for arbitrary radii). In these works the method for solid spheres is presented, that is, for coefficients ${{A}_{1}},{{A}_{2}},B$. We propose a development of this method for the case of variable radii, i.e. for the other seven coefficients.

We rewrite the coefficient  ${{A}_{1}}=\frac{R_{1}^{3}}{6}+{{c}^{3}}\sum\limits_{n=2}^{\infty }{\frac{{{e}^{-\left( 2n-1 \right){{\tau }_{1}}}}}{{{e}^{\left( 2n-1 \right)\left( {{\tau }_{1}}+{{\tau }_{2}} \right)}}-1}}\frac{{{(2n-1)}^{2}}-1}{2}$ as follows
\begin{equation}\label{A1forMellin}
  {{A}_{1}}=\frac{R_{1}^{3}}{6}+{{c}^{3}}\sum\limits_{n=2}^{\infty }{\frac{{{e}^{-\left( 1+{{\lambda }_{1}} \right)t}}}{1-{{e}^{-t}}}}\frac{{{(2n-1)}^{2}}-1}{2},~~~~~~t=\left( 2n-1 \right)\left( {{\tau }_{1}}+{{\tau }_{2}} \right),~~~~~{{\lambda }_{1}}=\frac{{{\tau }_{1}}}{{{\tau }_{1}}+{{\tau }_{2}}}.
\end{equation}
Substituting under the sign of sum the Mellin transform
\begin{equation}\label{Mellin}
\frac{{{e}^{-(1+{{\lambda }_{1}})t}}}{1-{{e}^{-t}}}=\frac{1}{2\pi i}\int\limits_{\sigma -i\infty }^{\sigma +i\infty }{{{t}^{-s}}\Gamma (s)}\zeta \left( s,1+{{\lambda }_{1}} \right)ds,\sigma >1,\,\operatorname{Re}\,(1+{{\lambda }_{1}})>0,
\end{equation}
where  $\zeta (s,a)=\sum\limits_{n=0}^{\infty }{\frac{1}{{{(n+a)}^{s}}}}$  is the Hurwitz zeta function, $\Gamma(s)$ is the gamma function, a new expression is obtained

\begin{equation}\label{A1Mellin}
\begin{aligned}
 {{A}_{1}}=&\frac{R_{1}^{3}}{6}+\frac{{{c}^{3}}}{2}\sum\limits_{n=2}^{\infty }{\frac{1}{2\pi i}\int\limits_{\sigma -i\infty }^{\sigma +i\infty }{{{\left( \left( 2n-1 \right)\left( {{\tau }_{1}}+{{\tau }_{2}} \right) \right)}^{-s}}\left( {{(2n-1)}^{2}}-1 \right)\Gamma (s)}}\zeta \left( s,1+{{\lambda }_{1}} \right)ds \\
  =&\frac{R_{1}^{3}}{6}+\frac{{{c}^{3}}}{2}\frac{1}{2\pi i}\int\limits_{\sigma -i\infty }^{\sigma +i\infty }{{{\left( {{\tau }_{1}}+{{\tau }_{2}} \right)}^{-s}}\Gamma (s)}Z\left( s \right)\zeta \left( s,1+{{\lambda }_{1}} \right)ds,\quad \sigma >3 ,\\
\end{aligned}
\end{equation}
where
\begin{equation}\label{Zs}
Z(s)=\sum\limits_{n=2}^{\infty }{{{\left( 2n-1 \right)}^{-s}}\left( {{(2n-1)}^{2}}-1 \right)}=\zeta (s-2)(1-{{2}^{-(s-2)}})-\zeta (s)(1-{{2}^{-s}}),
\end{equation}
$\zeta (s)=\zeta (s,1)$  is the Riemann zeta function.

This integral is calculated by using the residue theorem. We should find the poles of the integrated function. They are located in points $3,1,-1,\ldots $ and determine the order of the asymptotic expansion terms. The residue in the first point determines the main expansion coefficient, in the second point - the next one, etc. Considering the residues in $3,1,-1,\ldots,-2l+1$, we obtain that \cite{raszillier1990optimal}
\begin{equation}\label{A1asympt}
\begin{aligned}
{{A}_{1}}=&\frac{R_{1}^{3}}{6}+\frac{{{c}^{3}}}{2}\left( \frac{\zeta (3,1+{{\lambda }_{1}})}{{{({{\tau }_{1}}+{{\tau }_{2}})}^{3}}}+\frac{1}{2}\frac{1}{{{\tau }_{1}}+{{\tau }_{2}}}\ln \frac{{{\tau }_{1}}+{{\tau }_{2}}}{2}+\frac{1}{2}\frac{1}{{{\tau }_{1}}+{{\tau }_{2}}}\left( \psi (1+{{\lambda }_{1}})+\frac{1}{6} \right) \right. \\
& \left. -\sum\limits_{k=1}^{l}{\frac{{{({{\tau }_{1}}+{{\tau }_{2}})}^{2k-1}}}{(2k-1)!}Z(-2k+1)}\,\zeta (-2k+1,1+{{\lambda }_{1}}) \right)+r_{{{A}_{1}}}^{2l-1}, \\
\end{aligned}
\end{equation}
where $\psi(x)=\Gamma'(x)/\Gamma(x)$ is the digamma function, the residue term is
\begin{equation}\label{restA1asympt}
r_{{{A}_{1}}}^{m}=\frac{{{c}^{3}}}{2}\frac{1}{2\pi i}\int\limits_{{{\sigma }_{m}}-i\infty }^{{{\sigma }_{m}}+i\infty }{{{\left( {{\tau }_{1}}+{{\tau }_{2}} \right)}^{-s}}\Gamma (s)}Z\left( s \right)\zeta \left( s,1+{{\lambda }_{1}} \right)ds,~~~-m-1<{{\sigma }_{m}}<-m.
\end{equation}
Similarly, for $B$  Raszillier \emph{et al.} \cite{raszillier1990optimal} obtained

\begin{equation}\label{Basympt}
\begin{aligned}
   B=&\frac{{{c}^{3}}}{2}\frac{1}{2\pi i}\int\limits_{\sigma -i\infty }^{\sigma +i\infty }{{{\left( {{\tau }_{1}}+{{\tau }_{2}} \right)}^{-s}}\Gamma (s)}Z\left( s \right)\zeta \left( s \right)ds \\
  =&\frac{{{c}^{3}}}{2}\left( \frac{\zeta (3)}{{{({{\tau }_{1}}+{{\tau }_{2}})}^{3}}}+\frac{1}{2}\frac{1}{{{\tau }_{1}}+{{\tau }_{2}}}\ln \frac{{{\tau }_{1}}+{{\tau }_{2}}}{2}+\frac{1}{2}\frac{1}{{{\tau }_{1}}+{{\tau }_{2}}}\left( \psi (1)+\frac{1}{6} \right) \right. \\
 & \left. -\sum\limits_{k=1}^{n}{\frac{{{({{\tau }_{1}}+{{\tau }_{2}})}^{2k-1}}}{(2k-1)!}Z(-2k+1)}\,\zeta (-2k+1) \right)+r_{B}^{2l-1}, \\
\end{aligned}
\end{equation}
where  $r_{B}^{2l-1}$ is defined analogically to $r_{{{A}_{1}}}^{2l-1}$.

Let us provide the asymptotic expansion of the other coefficients

\begin{equation}\label{C11asympt}
\begin{aligned}
    {{C}_{11}}=&\frac{2{{c}^{3}}}{{{\sinh }^{2}}{{\tau }_{1}}}\frac{1}{2\pi i}\int\limits_{\sigma -i\infty }^{\sigma +i\infty }{{{\left( {{\tau }_{1}}+{{\tau }_{2}} \right)}^{-s}}\Gamma (s)}{{Z}_{C_{11}}}(s,{{\lambda }_{1}})ds \\
  =&\frac{2{{c}^{3}}}{{{\sinh }^{2}}{{\tau }_{1}}}\left( -\frac{\sinh {{\tau }_{1}}}{2{{({{\tau }_{1}}+{{\tau }_{2}})}^{2}}}\zeta (2,1+{{\lambda }_{1}}) \right.+\frac{\cosh {{\tau }_{1}}-1}{{{\tau }_{1}}+{{\tau }_{2}}}\left( 1+\frac{1}{2}\ln \frac{{{\tau }_{1}}+{{\tau }_{2}}}{2} \right) \\
 & \left.+\frac{\gamma +\cosh {{\tau }_{1}}\,\psi (1+{{\lambda }_{1}})}{2({{\tau }_{1}}+{{\tau }_{2}})} +\frac{\sinh {{\tau }_{1}}}{{{\tau }_{1}}+{{\tau }_{2}}}+\sum\limits_{k=0}^{m}{{{(-1)}^{k}}\frac{{{({{\tau }_{1}}+{{\tau }_{2}})}^{k}}}{k!}{{Z}_{C_{11}}}(-k,{{\lambda }_{1}})} \right)+r_{{{C}_{11}}}^{m}, \\
\end{aligned}
\end{equation}
\begin{equation}\label{C12asympt}
\begin{aligned}
   {{C}_{12}}=&\frac{2{{c}^{3}}}{{{\sinh }^{2}}{{\tau }_{2}}}\frac{1}{2\pi i}\int\limits_{\sigma -i\infty }^{\sigma +i\infty }{{{\left( {{\tau }_{1}}+{{\tau }_{2}} \right)}^{-s}}\Gamma (s)}{{Z}_{C_{12}}}(s,{{\lambda }_{1}})ds \\
  =&\frac{2{{c}^{3}}}{{{\sinh }^{2}}{{\tau }_{2}}}\left( -\frac{\sinh {{\tau }_{2}}}{2{{({{\tau }_{1}}+{{\tau }_{2}})}^{2}}}\zeta (2) \right.+\frac{\cosh {{\tau }_{2}}-1}{{{\tau }_{1}}+{{\tau }_{2}}}\left( 1+\frac{1}{2}\ln \frac{{{\tau }_{1}}+{{\tau }_{2}}}{2} \right)- \\
 & \left. - \frac{\gamma \cosh {{\tau }_{2}}\,+\psi ({{\lambda }_{1}})}{2({{\tau }_{1}}+{{\tau }_{2}})}+\frac{\sinh {{\tau }_{2}}}{{{\tau }_{1}}+{{\tau }_{2}}}+\sum\limits_{k=0}^{m}{{{(-1)}^{k}}\frac{{{({{\tau }_{1}}+{{\tau }_{2}})}^{k}}}{k!}{{Z}_{C_{12}}}(-k,{{\lambda }_{1}})} \right)+r_{{{C}_{12}}}^{m}, \\
\end{aligned}
\end{equation}
where $\gamma$  is the Euler--Mascheroni constant,
\begin{equation}\label{ZC11ZC12}
\begin{aligned}
 {{Z}_{{{C}_{11}}}}(s,{{\lambda }_{1}})=&\left( \zeta (s)-\zeta (s,1+{{\lambda }_{1}})\cosh {{\tau }_{1}} \right)\left( \zeta (s)(1-{{2}^{-s}})-1 \right)\\
&-\zeta (s,1+{{\lambda }_{1}})\sinh {{\tau }_{1}}\left( \zeta (s-1)(1-{{2}^{-(s-1)}})-1 \right), \\
{{Z}_{{{C}_{12}}}}(s,{{\lambda }_{1}})=&\left( \zeta (s,{{\lambda }_{1}})-\zeta (s)\cosh {{\tau }_{2}} \right)\left( \zeta (s)(1-{{2}^{-s}})-1 \right)\\
&-\zeta (s)\sinh {{\tau }_{2}}\left( \zeta (s-1)(1-{{2}^{-(s-1)}})-1 \right). \\
\end{aligned}
\end{equation}

For coefficients ${{D}_{1}},E$ the proof of the asymptotic expansion is much more complicated. It may be found in Appendix \ref{AppendixAsymptDE}. ${{A}_{2}},{{C}_{22}},{{C}_{21}},{{D}_{2}}$ can be found using subscript permutation.

\subsection{Comparison of the asymptotic expansion}
Raszillier et.al. \cite{raszillier1990optimal} compared the asymptotic expansion of kinetic energy with the three term expansion, obtained by Voinov \cite{Voinov1969PMM} for two spheres of constant radii and proved that they fully coincide. One may also compare the asymptotic expansions of kinetic energy, obtained above with the three term expansion from \cite{sanduleanu2018trinomial}. They fully coincide  (see Appendix \ref{AppendixEqualDAN}).

\subsection{Estimation of the residue term   }
By Poincare \cite{whittaker1996course}, the divergent series $X=\sum\limits_{n=0}^{m}{{{X}_{n}}}(\varepsilon )+R_{X}^{m}(\varepsilon )$    is said to be an asymptotic expansion if
$\underset{\varepsilon \to 0}{\mathop{\lim }}\,\left| \frac{R_{X}^{m}}{{{X}_{m}}} \right|=0$.
 	
For example, let us consider the expansion of     ${{A}_{1}}$ (\ref{A1asympt}). We present it as follows

\begin{equation}\label{A1asymptseries}
{{A}_{1}}=x_{{{A}_{1}}}^{0}+\sum\limits_{k=1}^{l}{x_{{{A}_{1}}}^{2k-1}}+r_{{{A}_{1}}}^{2l-1},
\end{equation}
where
\begin{equation}\label{xA1asymptseries}
x_{{{A}_{1}}}^{2k-1}=-\frac{R_{1}^{3}{{\sinh }^{3}}{{\tau }_{1}}}{2}\frac{{{({{\tau }_{1}}+{{\tau }_{2}})}^{2k-1}}}{(2k-1)!}Z(-2k+1)\zeta (-2k+1,1+{{\lambda }_{1}}).
\end{equation}

Let us prove that  $\underset{{{\tau }_{1}}+{{\tau }_{2}}\to 0}{\mathop{\lim }}\,\left| \frac{r_{{{A}_{1}}}^{2l-1}}{x_{{{A}_{1}}}^{2l-1}} \right|=0$.

For the expression
\begin{equation}\label{rA1abs}
\begin{aligned}
  & \left| r_{{{A}_{1}}}^{2l-1} \right|=\frac{R_{1}^{3}{{\sinh }^{3}}{{\tau }_{1}}}{2}\frac{1}{2\pi }{{\left( {{\tau }_{1}}+{{\tau }_{2}} \right)}^{-{{\sigma }_{2l-1}}}}\int\limits_{-\infty }^{\infty }{\left| \Gamma ({{\sigma }_{2l-1}}+it)Z\left( {{\sigma }_{2l-1}}+it \right)\zeta \left( {{\sigma }_{2l-1}}+it,1+{{\lambda }_{1}} \right) \right|}dt, \\
 & -2l<{{\sigma }_{2l-1}}<-2l+1, \\
\end{aligned}
\end{equation}
in \cite{raszillier1990optimal} the following estimate was obtained
\begin{equation}\label{rA1O}
\left| r_{{{A}_{1}}}^{2l-1} \right|=O\left( {{({{\tau }_{1}}+{{\tau }_{2}})}^{2l-1+3}} \right).
\end{equation}
Also note that
\begin{equation}\label{xA1O}
\left| x_{{{A}_{1}}}^{2l-1} \right|=O\left( {{({{\tau }_{1}}+{{\tau }_{2}})}^{2l-1+3}} \right).
\end{equation}
It turns out that one may also prove that
\begin{equation}\label{rA1otoprove}
\left| r_{{{A}_{1}}}^{2l-1} \right|=o\left( {{({{\tau }_{1}}+{{\tau }_{2}})}^{2l-1+3}} \right).
\end{equation}
We substitute
\begin{equation}\label{rA1o}
\left| r_{{{A}_{1}}}^{2l-1} \right|=\left| x_{{{A}_{1}}}^{2l+1}+r_{{{A}_{1}}}^{2l+1} \right|\le \left| x_{{{A}_{1}}}^{2l+1} \right|+\left| r_{{{A}_{1}}}^{2l+1} \right|=O\left( {{({{\tau }_{1}}+{{\tau }_{2}})}^{2l+1+3}} \right)=o\left( {{({{\tau }_{1}}+{{\tau }_{2}})}^{2l-1+3}} \right)
\end{equation}
and  obtain that
\begin{equation}\label{rA1xA1}
\underset{{{\tau }_{1}}+{{\tau }_{2}}\to 0}{\mathop{\lim }}\,\left| \frac{r_{{{A}_{1}}}^{2l-1}}{x_{{{A}_{1}}}^{2l-1}} \right|=\underset{{{\tau }_{1}}+{{\tau }_{2}}\to 0}{\mathop{\lim }}\,\frac{o\left( {{({{\tau }_{1}}+{{\tau }_{2}})}^{2l-1+3}} \right)}{O\left( {{({{\tau }_{1}}+{{\tau }_{2}})}^{2l-1+3}} \right)}=0,
\end{equation}
and, thus, prove that the expansion of ${{A}_{1}}$ is asymptotic. The series diverges for any ${{{\tau }_{1}}+{{\tau }_{2}}>0}$ and for numerical calculations we must consider only a finite number of series terms.

As noted in \cite{dingle1973asymptotic}, it is reasonable to truncate the sum of the asymptotic series  at   ${m=\eta} $, where $\eta $  can be found from the following equation   ${{\left. \frac{d\left| x_{{{A}_{1}}}^{m} \right|}{dm} \right|}_{\eta }}\sim 0$.

Let us estimate the dependence of $x_{{{A}_{1}}}^{m}$ for large values of $m$. Taking into consideration the equality
\begin{equation}\label{Zetaaplus1}
\zeta (-m,a+1)=\zeta (-m,a)+{{a}^{m}}
\end{equation}
and Hurwitz's formula  \cite{apostol2013}
\begin{equation}\label{Hurwitzformula}
\zeta (1-m,a)=2\frac{(m-1)!}{{{(2\pi )}^{m}}}\sum\limits_{n=1}^{\infty }{\frac{\cos (2\pi na-\frac{1}{2}\pi m)}{{{n}^{m}}}},\quad 1\ge a\ge 0, \quad m\ge 1,
\end{equation}
$\zeta (-m,a)$ may be approximated for large $m$ and for $0\le a\le 2$ as follows
\begin{equation}\label{Zetaapprox}
\zeta (-m,a)=2\frac{m!}{{{(2\pi )}^{m+1}}}\cos \left( 2\pi a-\frac{1}{2}\pi (m+1) \right),
\end{equation}
and, thus, $x_{{{A}_{1}}}^{m}$ may be estimated as
\begin{equation}\label{xA1estimation}
\begin{aligned}
   \frac{x_{{{A}_{1}}}^{m}}{{{p}^{3}}}\sim & {{({{\tau }_{1}}+{{\tau }_{2}})}^{3}}\frac{{{({{\tau }_{1}}+{{\tau }_{2}})}^{m}}}{m!}2\frac{(m+2)!}{{{(2\pi )}^{m+3}}}{{2}^{m+2}}2\frac{m!}{{{(2\pi )}^{m+1}}}\sim 4({{\tau }_{1}}+{{\tau }_{2}})\frac{{{(2({{\tau }_{1}}+{{\tau }_{2}}))}^{m+2}}(m+2)!}{{{(2\pi )}^{2m+4}}} \\
 \sim & 4({{\tau }_{1}}+{{\tau }_{2}})\sqrt{2\pi (m+2)}{{\left( \frac{({{\tau }_{1}}+{{\tau }_{2}})(m+2)}{2{{\pi }^{2}}e} \right)}^{(m+2)}}. \\
\end{aligned}
\end{equation}

Recall that  $p=\frac{{{R}_{1}}{{R}_{2}}}{{{R}_{1}}+{{R}_{2}}}$.
The condition  ${{\left. \frac{d\left| x_{{{A}_{1}}}^{m} \right|}{dm} \right|}_{\eta }} \sim 0$ implies that  ${\eta \sim \frac{2{{\pi }^{2}}}{{{\tau }_{1}}+{{\tau }_{2}}}}$. Considering formula (\ref{t1plust2}), at small separation distance we obtain that  $\eta \sim \frac{2{{\pi }^{2}}}{\sqrt{2h/p}}$. Analogically, we calculate the values of $\eta $  for all the other coefficients. With such choice of $\eta $ the error is of order ${{e}^{-\eta }}$. This estimate is confirmed by numerous numerical calculations.	

\subsection{Expansion in $h$  at small separation distance}
In practice it is more convenient to use instead of parameter ${{\tau }_{1}}+{{\tau }_{2}}$ the separation distance $h$. Then the kinetic energy coefficients' expansion is
\begin{equation}\label{Xbyh}
X={{f}_{X}}(h)+{{g}_{X}}(h)\ln \left( \frac{h}{2p} \right),~~~ p=\frac{{{R}_{1}}{{R}_{2}}}{{{R}_{1}}+{{R}_{2}}},~~~X=\{{{A}_{i}},B,{{C}_{ij}},{{D}_{i}},E\}.
\end{equation}
We need to find 6 pairs of functions  ${{f}_{X}}(h)$ and  ${{g}_{X}}(h)$  for the coefficients ${{A}_{1}}$,$B$,${{C}_{11}}$,${{C}_{12}}$, ${{D}_{1}}$,$E$, thus, in total, 12 functions. For the rest of the coefficients functions ${{f}_{X}}(h)$ and ${{g}_{X}}(h)$  are obtained by subscript permutation.

Note that the number of independent functions may be reduced to 10. To support this statement, let us prove that functions  ${{g}_{X}}(h)$ coincide for coefficients $X=\{ {{A}_{1}},B\}$ and for coefficients  $X=\{{{C}_{11}},{{C}_{21}}\}$.

Indeed, (\ref{A1asympt}) and (\ref{Basympt}) imply that coefficients ${{g}_{{{A}_{1}}}},{{g}_{B}}$  are obtained from the term
\begin{equation}\label{gA1Bgener}
\frac{{{c}^{3}}}{2}\frac{1}{2}\frac{1}{{{\tau }_{1}}+{{\tau }_{2}}}\ln \frac{{{\tau }_{1}}+{{\tau }_{2}}}{2},
\end{equation}
where ${{\tau }_{1}}+{{\tau }_{2}}$ should be expressed through $h$. Coefficients ${{g}_{{{C}_{11}}}},{{g}_{{{C}_{21}}}}$ are obtained similarly from
\begin{equation}\label{gC11C21gener}
\frac{{{c}^{3}}}{{{\sinh }^{2}}{{\tau }_{1}}}\frac{\cosh {{\tau }_{1}}-1}{{{\tau }_{1}}+{{\tau }_{2}}}\ln \frac{{{\tau }_{1}}+{{\tau }_{2}}}{2}.
\end{equation}

Functions ${{f}_{X}}(h)$ and ${{g}_{X}}(h)$ can be expanded by powers of $h$. Sufficient accuracy is achieved by the following cubic polynomial
\begin{equation}\label{fhgh}
\begin{aligned}
  & {{f}_{X}}(h)=f_{X}^{0}+f_{X}^{1}h+f_{X}^{2}{{h}^{2}}+f_{X}^{3}{{h}^{3}}+O\left( {{h}^{4}} \right) \\
  & {{g}_{X}}(h)=g_{X}^{1}h+g_{X}^{2}{{h}^{2}}+g_{X}^{3}{{h}^{3}}+O\left( {{h}^{4}} \right) \\
\end{aligned}
\end{equation}
One can show that up to subscript permutation, the logarithmic singularity is determined by four polynomials. The first three coefficients of these polynomials are presented in table~\ref{table1}.


\begin{table}
\caption{Analytical form of $g_{X}^{1},g_{X}^{2},g_{X}^{3}$ for the kinetic energy coefficients}
\label{table1}
\begin{center}
\begin{tabular}{|l|l|l|l|}
\hline
$X$ &	$g_{X}^{1}$&	$g_{X}^{2}$	&$g_{X}^{3}$\\
\hline
${{A}_{1}},B$ &	$\frac{{{p}^{2}}}{4}$ &	$\frac{5p}{24}\left( \alpha _{1}^{2}-{{\alpha }_{1}}{{\alpha }_{2}}+\alpha _{2}^{2} \right)$ &	$\frac{13\alpha _{1}^{4}-98\alpha _{1}^{3}{{\alpha }_{2}}+183\alpha _{2}^{2}\alpha _{2}^{2}-98{{\alpha }_{1}}\alpha _{2}^{3}+13\alpha _{2}^{4}}{360}$\\
\hline
${{C}_{11}},{{C}_{21}}$ &	$\frac{{{p}^{2}}}{2}$	&$\frac{p}{12}\left( 5\alpha _{1}^{2}-5{{\alpha }_{1}}{{\alpha }_{2}}+2\alpha _{2}^{2} \right)$ 	& $\frac{13\alpha _{1}^{4}-98\alpha _{1}^{3}{{\alpha }_{2}}+123\alpha _{2}^{2}\alpha _{2}^{2}-38{{\alpha }_{1}}\alpha _{2}^{3}-2\alpha _{2}^{4}}{180}$\\
\hline
${{D}_{1}}$ &	$\frac{{{p}^{2}}}{4}$	& $\frac{p}{24}\left( 5\alpha _{1}^{2}-5{{\alpha }_{1}}{{\alpha }_{2}}-\alpha _{2}^{2} \right)$ & 	$\frac{11}{720}+\frac{\alpha _{1}^{2}\left( \alpha _{1}^{2}-16{{\alpha }_{1}}{{\alpha }_{2}}+4\alpha _{2}^{2} \right)}{48}$\\
\hline
$E$ &	$\frac{{{p}^{2}}}{4}$ &	$\frac{p}{24}\left( -2\alpha _{1}^{2}+5{{\alpha }_{1}}{{\alpha }_{2}}-2\alpha _{2}^{2} \right)$ & 	$\frac{11}{720}-\frac{\left( \alpha _{1}^{2}-{{\alpha }_{1}}{{\alpha }_{2}}+\alpha _{2}^{2} \right)\left( \alpha _{1}^{2}+9{{\alpha }_{1}}{{\alpha }_{2}}+\alpha _{2}^{2} \right)}{48}$ \\
\hline
\end{tabular}
\end{center}
\end{table}
The polynomials ${{f}_{X}}(h)$ are more lengthy. Thus, it is more convenient to give the numerical values of coefficients of polynomials ${{f}_{X}}(h)$ for given radii ratio. They are given in tables \ref{table2}-\ref{table4} for radii ratio ${{R}_{2}}/{{R}_{1}}=\{1,3,10\}$ accordingly.

\begin{table}
\caption{ Numerical values $f_{X}^{0},f_{X}^{1},f_{X}^{2},f_{X}^{3}$  for the kinetic energy coefficients at $R_2/R_1=1$}
\label{table2}
\begin{center}
\begin{tabular}{ | l | l | l | l | l | }
\hline
	$X$ & $f_X^0/R_1^3$ & $f_X^1/R_1^3$ & $f_X^2/R_1^3$ & $f_X^3/R_1^3$ \\ \hline
	$A_1$ & 0.19257 & 0.03834 & -0.05783 & -0.0064 \\ \hline
	$B$ & 0.07513 & -0.01375 & -0.03339 & -0.00841 \\ \hline
	$A_2$ & 0.19257 & 0.03834 & -0.05783 & -0.0064 \\ \hline
	$C_{11}$ & 0.07315 & 0.02403 & -0.09609 & 0.00345 \\ \hline
	$C_{12}$ & 0.28191 & -0.12419 & -0.04413 & -0.00127 \\ \hline
	$C_{21}$ & 0.28191 & -0.12419 & -0.04413 & -0.00127 \\ \hline
	$C_{22}$ & 0.07315 & 0.02403 & -0.09609 & 0.00345 \\ \hline
	$D_1$ & 1.05634 & -0.02539 & -0.02837 & 0.00549 \\ \hline
	$E$ & 0.52088 & -0.20799 & 0.02508 & -0.00506 \\ \hline
	$D_2$ & 1.05634 & -0.02539 & -0.02837 & 0.00549 \\ \hline
\end{tabular}
\end{center}
\end{table}

\begin{table}
\caption{ Numerical values $f_{X}^{0},f_{X}^{1},f_{X}^{2},f_{X}^{3}$  for the kinetic energy coefficients at $R_2/R_1=3$}
\label{table3}
\begin{center}
\begin{tabular}{ | l | l | l | l | l | }
\hline
	$X$ & $f_X^0/R_1^3$ & $f_X^1/R_1^3$ & $f_X^2/R_1^3$ & $f_X^3/R_1^3$ \\ \hline
	$A_1$ & 0.22593 & 0.15456 & -0.12096 & -0.00727 \\ \hline
	$B$ & 0.25356 & 0.03244 & -0.08352 & -0.00943 \\ \hline
	$A_2$ & 4.64004 & 0.05236 & -0.0598 & -0.02441 \\ \hline
	$C_{11}$ & 0.18871 & 0.14903 & -0.22995 & 0.04843 \\ \hline
	$C_{12}$ & 0.65753 & 0.07208 & -0.20445 & -0.00445 \\ \hline
	$C_{21}$ & 2273271 & -0.59585 & 0.00243 & -0.01 \\ \hline
	$C_{22}$ & 0.3404 & 0.06923 & -0.13721 & -0.03191 \\ \hline
	$D_1$ & 1.18701 & -0.07682 & -0.03699 & 0.01731 \\ \hline
	$E$ & 2.32151 & -0.45781 & -0.00579 & 0.00787 \\ \hline
	$D_2$ & 27.21124 & 0.019 & -0.08045 & -0.00683 \\ \hline
\end{tabular}
\end{center}
\end{table}

\begin{table}
\caption{ Numerical values $f_{X}^{0},f_{X}^{1},f_{X}^{2},f_{X}^{3}$  for the kinetic energy coefficients at $R_2/R_1=10$}
\label{table4}
\begin{center}
\begin{tabular}{ | l | l | l | l | l | }
\hline
	$X$ & $f_X^0/R_1^3$ & $f_X^1/R_1^3$ & $f_X^2/R_1^3$ & $f_X^3/R_1^3$ \\ \hline
	$A_1$ & 0.25175 & 0.29173 & -0.11741 & -0.04509 \\ \hline
	$B$ & 0.45156 & 0.20392 & -0.1117 & -0.04079 \\ \hline
	$A_2$ & 167.02413 & 0.16238 & -0.09517 & -0.03972 \\ \hline
	$C_{11}$ & 0.28907 & 0.33643 & -0.31326 & 0.03108 \\ \hline
	$C_{12}$ & 0.98412 & 0.45188 & -0.23408 & -0.08365 \\ \hline
	$C_{21}$ & 8.51296 & -1.12663 & -0.13156 & 0.01537 \\ \hline
	$C_{22}$ & 0.77253 & 0.34336 & -0.20516 & -0.07911 \\ \hline
	$D_1$ & 1.35988 & -0.10779 & -0.0565 & 0.02402 \\ \hline
	$E$ & 9.22154 & -0.64576 & -0.09002 & 0.012 \\ \hline
	$D_2$ & 1000.41854 & 0.18447 & -0.10969 & -0.03978 \\ \hline
\end{tabular}
\end{center}
\end{table}

\begin{figure}
\centering
\begin{minipage}{0.49\linewidth}
\centering
\includegraphics[width=1.\linewidth]{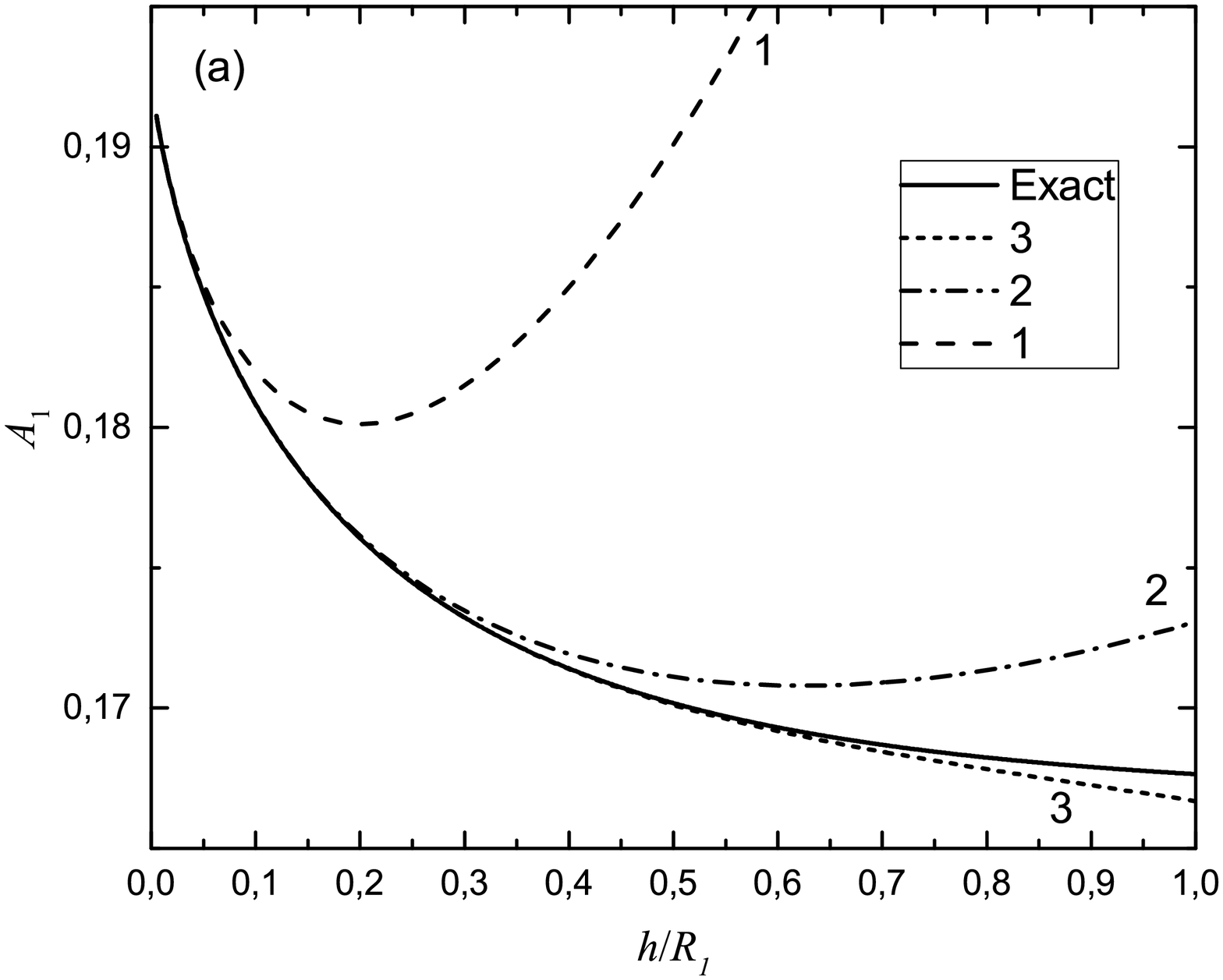}
\end{minipage}
\hfill
\begin{minipage}{0.49\linewidth}
\centering
\includegraphics[width=1.\linewidth]{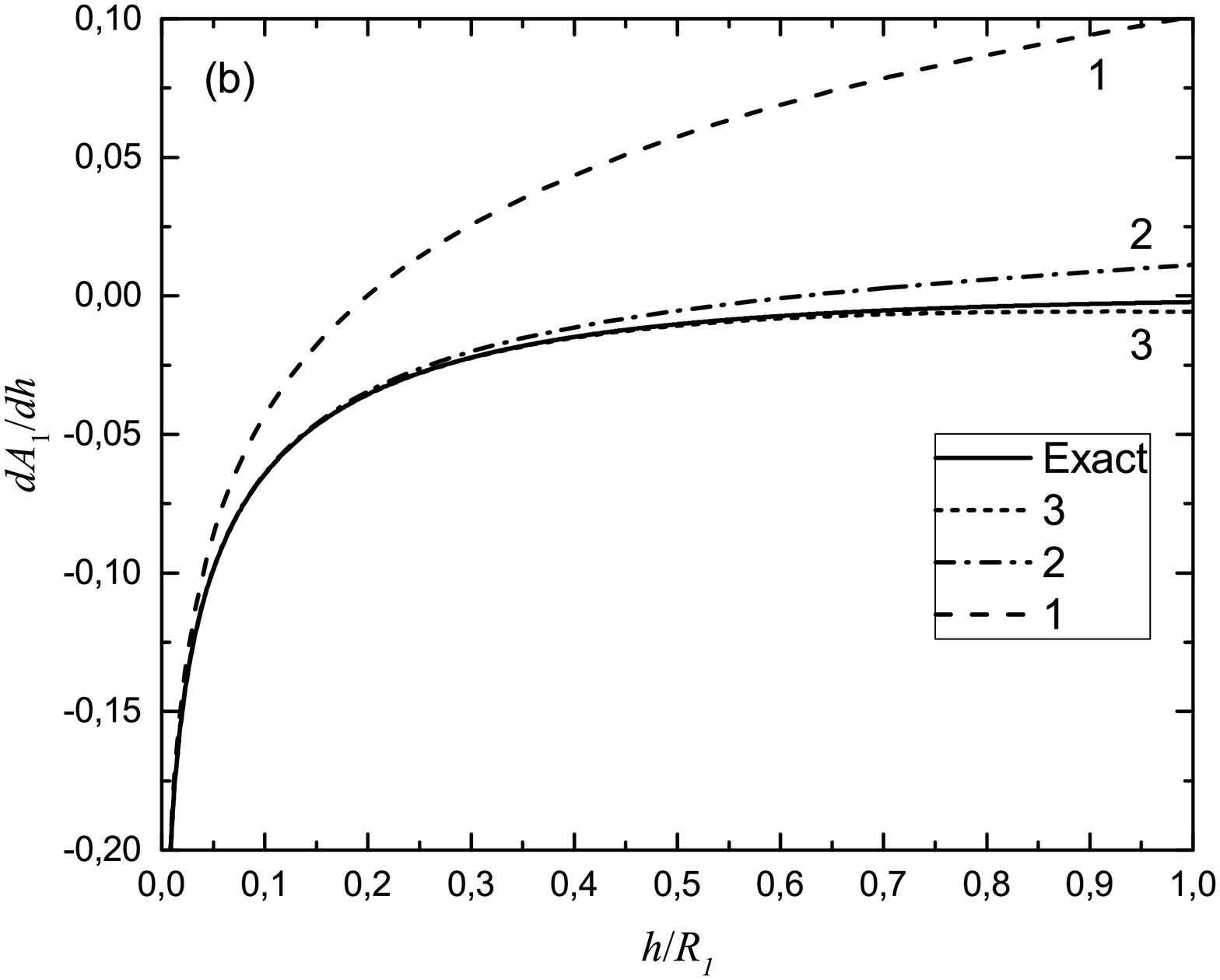}
\end{minipage}
\caption{Convergence of approximations (a) of coefficient  ${{A}_{1}}$ and (b) its derivative $d{{A}_{1}}/dh$  by first (dash), second (dash dot) and third degree (short dash) polynomials to exact dependences (solid).}
\label{fig2}
\end{figure}

\begin{figure}
\centering
\begin{minipage}[h]{0.48\linewidth}
\centering
\includegraphics[width=1.\linewidth]{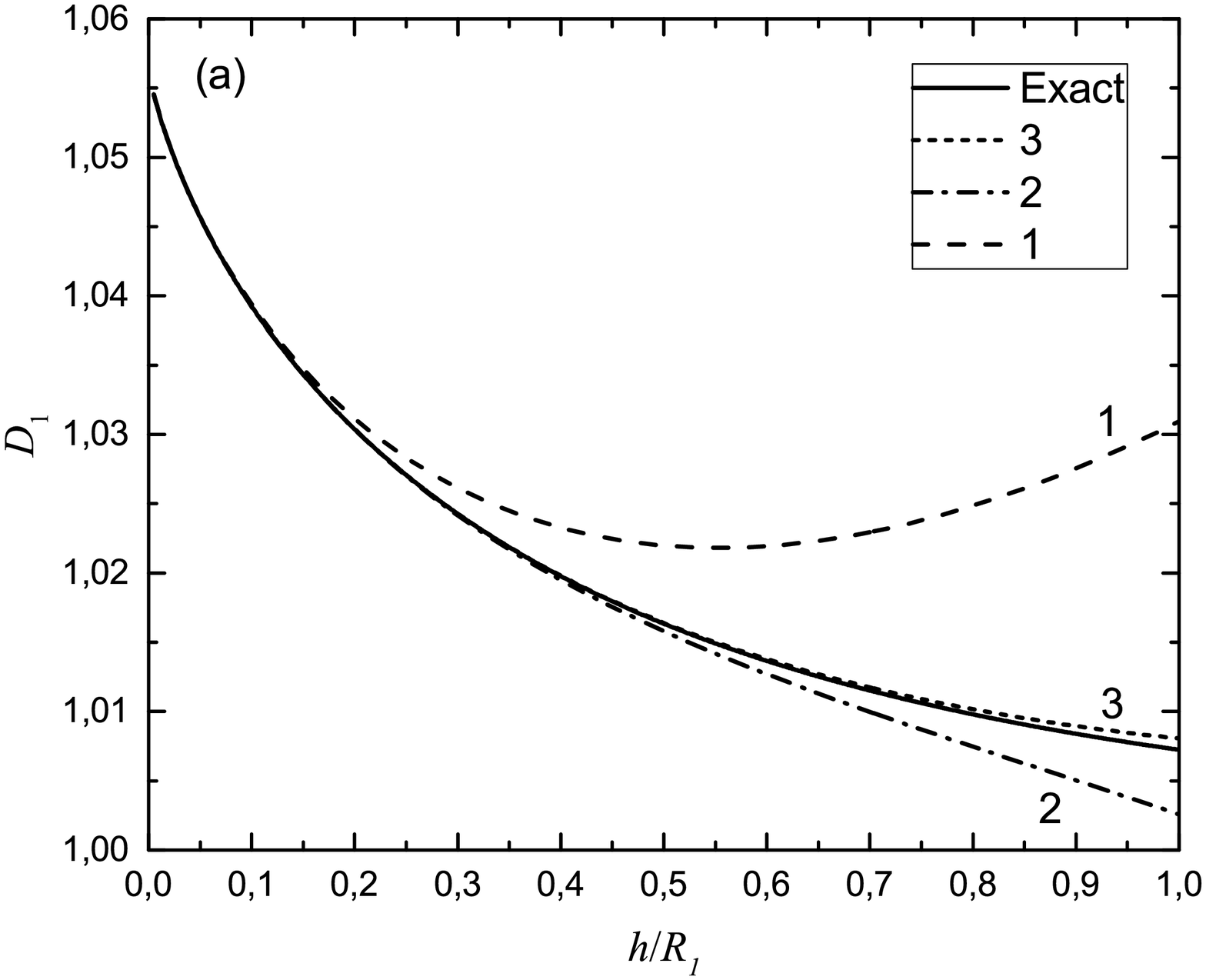}
\end{minipage}
\hfill
\begin{minipage}[h]{0.48\linewidth}
\centering
\includegraphics[width=1.\linewidth]{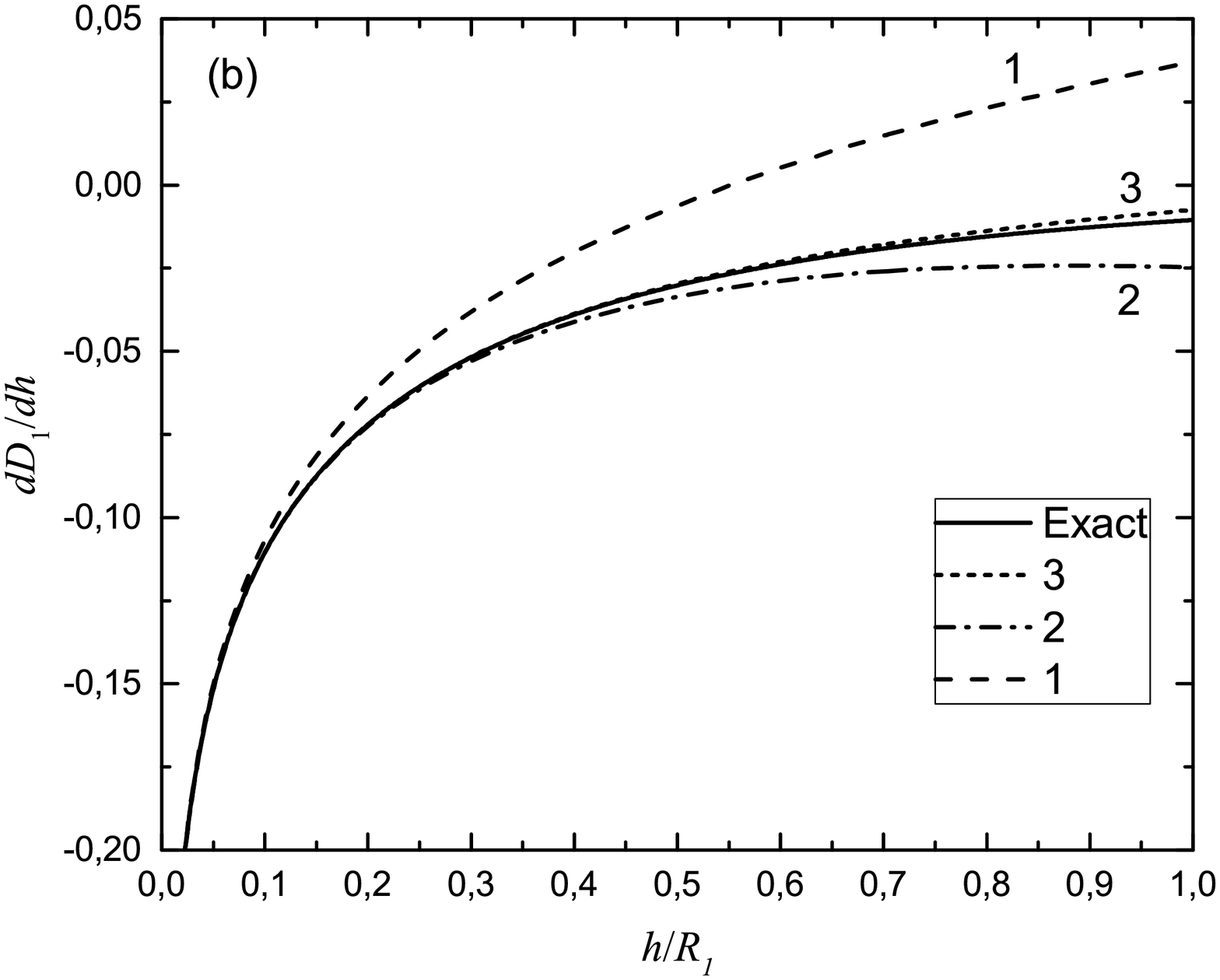}
\end{minipage}
\caption{Convergence of approximations (a) of coefficient  ${{D}_{1}}$ and (b) its derivative $d{{D}_{1}}/dh$  by first (dash), second (dash dot) and third degree (short dash) polynomials to exact dependences (solid).}
\label{fig3}
\end{figure}

In Figs. \ref{fig2}a and \ref{fig3}a is presented a comparison of coefficients ${{A}_{1}}$ and ${{D}_{1}}$ calculated exactly (solid lines) with their approximations by polynomials of first (dash), second(dash dot) and third (short dash) orders. For their derivatives a comparison is shown in Fig. \ref{fig2}b and \ref{fig3}b. As shown in the pictures, the increase of the polynomial degree gives a considerable increase of accuracy.

\subsection{Hydrodynamic force }
The hydrodynamic force, acting upon a sphere, is determined by the Lagrange formula
\begin{equation}\label{F1Lagr}
{{F}_{1}}=-\frac{d}{dt}\frac{\partial T}{\partial {{{\dot{z}}}_{1}}}+\frac{\partial T}{\partial {{z}_{1}}}=\frac{d}{dt}\frac{\partial T}{\partial {{u}_{1}}}+\frac{\partial T}{\partial r}=\frac{d}{dt}\frac{\partial T}{\partial {{u}_{1}}}+\frac{\partial T}{\partial h}.
\end{equation}
With the help of this formula and kinetic energy coefficients asymptotic expansions, one may get the expansion of the force at small separation distance with any accuracy by $h$.
The main hydrodynamic force asymptotic at small separation distance is \cite{petrov2013three}
\begin{equation}\label{F1asympt}
\frac{{{F}_{1}}}{2\pi \rho }=-\frac{d}{dt}\left( \frac{1}{2}{{p}^{2}}h\ln \left( \frac{h}{2p} \right)\dot{h} \right)+\frac{1}{4}{{p}^{2}}\ln \left( \frac{h}{2p} \right){{\dot{h}}^{2}},
\end{equation}
where $\dot{h}=-\,({{u}_{1}}+{{u}_{2}}+{{\dot{R}}_{1}}+{{\dot{R}}_{2}})$.
This logarithmic singularity can hardly be obtained if one presents the kinetic energy as a finite series by the inverse powers of the distance between the bubbles' centers $r$.
	
In \cite{witze1968flow} for a dilating sphere such that  $R=\beta {{t}^{1/2}}$, which is in contact with a plane, it was obtained that the attraction force is  $F=0.29\pi {{\beta }^{4}}{{\rho }_{l}}$.
This result agrees with the one obtained using asymptotic expansion, suggested in this work $F=0.288954\pi {{\beta }^{4}}{{\rho }_{l}}$.

\section{Conclusion}
The exact expression for the stream function was obtained for two spheres of variable radii moving in fluid. It generalizes the stream function approach used by Bentwich and Miloh for solid spheres.
Using the stream function found, a new formula for the fluid kinetic energy is derived, in which the coefficients of the quadratic form are presented as infinite series. It is shown that these series coincide with Hicks's and Voinov's series. The advantage of these new series is allowing the expansion by a small separation distance instead of the usually used distance between bubbles' centers.
Using the new formula for the kinetic energy, the asymptotic expansions of the kinetic energy coefficients at small separation distance are found. The expansion is proved to be accurate up to the exponentially small residue term.
The asymptotics found are necessary for describing the dynamics of spherical bubbles at small separation distance, and can be used for the analysis of their possible coalescence (for example, under acoustic influence)

\section*{ACKNOWLEDGMENTS}
The author thanks Prof. Alexander Petrov for useful remarks and fruitfull discussions.

\appendix

\section{Functions expansion in Gegenbauer polynomials\label{AppendixGranCond}}
To solve the boundary problem (\ref{StreamFunctionEqBisher}) and (\ref{GranEq}) , we suggest to expand the left-hand sides of  (\ref{GranCond1Gege}), (\ref{GranCond2Gege}) and (\ref{GranCond3Gege}) in Gegenbauer polynomials. We use the following definition of Gegenbauer polynomials through the generation function \cite{whittaker1996course}

\begin{equation}\label{GenerFunc}
\frac{1}{{{(1-2x\mu +{{x}^{2}})}^{\nu }}}=\sum\limits_{n=0}^{\infty }{{{x}^{n}}C_{n}^{\nu }(\mu )}
\end{equation}

Substituting  $\nu =-\frac{1}{2},x={{e}^{-\tau }},\tau >0$ , we obtain (\ref{GranCond1Gege})

\begin{equation}\label{GenerFunctau}
{{\left( \cosh \tau -\mu  \right)}^{1/2}}=\sum\limits_{n=0}^{\infty }{C_{n}^{-1/2}(\mu )\frac{\sqrt{2}}{2}e^{-(n-1/2)\tau}},
\end{equation}
Differentiating by $\tau $, we obtain the following expression for  (\ref{GranCond2Gege})

\begin{equation}\label{GenerFunctaudiff}
\frac{1}{{{\left( \cosh \tau -\mu  \right)}^{1/2}}}=\sum\limits_{n=0}^{\infty }{C_{n}^{-1/2}(\mu )\frac{-(n-1/2)\sqrt{2}}{\sinh \tau }e^{-(n-1/2)\tau}}.
\end{equation}

To obtain (\ref{GranCond3Gege}) , we substitute   $\nu =\frac{1}{2}$ in (\ref{GenerFunc}) and  differentiate by $\mu $
\begin{equation}\label{GenerFuncDiffmu}
\frac{x}{{{(1-2x\mu +{{x}^{2}})}^{3/2}}}=\sum\limits_{n=0}^{\infty }{{{x}^{n}}\frac{dC_{n}^{1/2}(\mu )}{d\mu }}.
\end{equation}

As  $C_{0}^{\nu }(\mu )=1$ , we convert  (\ref{GenerFuncDiffmu}) into
\begin{equation}\label{App3/2}
\frac{1}{{{(1-2x\mu +{{x}^{2}})}^{3/2}}}=\sum\limits_{n=2}^{\infty }{{{x}^{n-2}}\frac{dC_{n-1}^{1/2}(\mu )}{d\mu }}.
\end{equation}

We need to express  $\frac{dC_{n-1}^{1/2}(\mu )}{d\mu }$ through $C_{n}^{-1/2}(\mu )$. As the following equality
\begin{equation}\label{GegeDiffmu}
\frac{dC_{n}^{\nu }(\mu )}{d\mu }=2\nu C_{n-1}^{\nu +1}(\mu )
\end{equation}
holds, for $\nu =-\frac{1}{2}$  we obtain
\begin{equation}\label{App3/2Diff0}
\frac{{{d}^{2}}C_{n}^{-1/2}(\mu )}{d{{\mu }^{2}}}=-\frac{dC_{n-1}^{1/2}(\mu )}{d\mu }.
\end{equation}

Further, taking into account the Gegenbauer differential equation \cite{whittaker1996course}
\begin{equation}\label{GegeDiffeq}
(1-{{\mu }^{2}})\frac{{{d}^{2}}C_{n}^{\nu }(\mu )}{d{{\mu }^{2}}}-(2\nu +1)\mu \frac{dC_{n}^{\nu }(\mu )}{d\mu }+n(n+2\nu )C_{n}^{\nu }(\mu )=0,
\end{equation}

for $\nu =-\frac{1}{2}$ we get that
\begin{equation}\label{App3/2Diff1}
(1-{{\mu }^{2}})\frac{{{d}^{2}}C_{n}^{-1/2}(\mu )}{d{{\mu }^{2}}}+n(n-1)C_{n}^{-1/2}(\mu )=0.
\end{equation}

Multiplying (\ref{App3/2}) by $1-{{\mu }^{2}}$ and considering (\ref{App3/2Diff0}) and (\ref{App3/2Diff1}), we obtain
\begin{equation}\label{App3/2Diff2}
\frac{1-{{\mu }^{2}}}{{{(1-2x\mu +{{x}^{2}})}^{3/2}}}=\sum\limits_{n=2}^{\infty }{{{x}^{n-2}}n(n-1)C_{n}^{-1/2}(\mu )}.
\end{equation}
From hereon, we start to sum from $0$ instead of $2$, as for $n=0$ and $n=1$ the elements of the sum equal zero. Substituting $x={{e}^{-\tau }},\tau >0$, we obtain  (\ref{GranCond3Gege})
\begin{equation}\label{App3/2Diff3}
\frac{1}{2}\frac{1-{{\mu }^{2}}}{{{\left( \cosh \tau -\mu  \right)}^{3/2}}}=\sum\limits_{n=0}^{\infty }{C_{n}^{-1/2}(\mu )\sqrt{2}n(n-1)e^{-(n-1/2)\tau}}.
\end{equation}

\section{Integrals \label{AppendixI1I6}}
To calculate the second integral in (\ref{KineticIntegrContour})  , we substitute in function $\psi $ the boundary conditions (\ref{GranValueStreamFunc}), $\psi {{'}_{\xi }}$ is found from (\ref{StreamFunctionSol}). We expand the brackets and get six integrals, for which if  $\xi >0$, $n\ge 2$ the following expressions  hold
\begin{equation}\label{IntI1I6}
\begin{aligned}
{{I}_{1}}=&\int\limits_{-1}^{1}{\frac{1-{{\mu }^{2}}}{{{\left( \cosh \xi -\mu  \right)}^{2}}}\frac{\cosh \xi -\mu }{1-{{\mu }^{2}}}}\frac{1}{{{\left( \cosh \xi -\mu  \right)}^{1/2}}}C_{n}^{-1/2}(\mu )d\mu =\frac{2\sqrt{2}}{n-1/2}{{e}^{-(n-1/2)\xi }} \\
{{I}_{2}}=&\int\limits_{-1}^{1}{\frac{1-{{\mu }^{2}}}{{{\left( \cosh \xi -\mu  \right)}^{2}}}\frac{\cosh \xi -\mu }{1-{{\mu }^{2}}}}\frac{\sinh \xi }{{{\left( \cosh \xi -\mu  \right)}^{3/2}}}C_{n}^{-1/2}(\mu )d\mu =\frac{1}{-3/2}\frac{d{{I}_{1}}}{d\xi }=\frac{4\sqrt{2}}{3}{{e}^{-(n-1/2)\xi }} \\
{{I}_{3}}=&\int\limits_{-1}^{1}{\frac{1}{\cosh \xi -\mu }\frac{\cosh \xi -\mu }{1-{{\mu }^{2}}}}\frac{1}{{{\left( \cosh \xi -\mu  \right)}^{1/2}}}C_{n}^{-1/2}(\mu )d\mu =\frac{\sqrt{2}}{n(n-1)}\frac{{{e}^{{{(-1)}^{n}}\xi /2}}-{{e}^{-(n-1/2)\xi }}}{\sinh \xi } \\
{{I}_{4}}=&\int\limits_{-1}^{1}{\frac{1}{\cosh \xi -\mu }\frac{\cosh \xi -\mu }{1-{{\mu }^{2}}}}\frac{\sinh \xi }{{{\left( \cosh \xi -\mu  \right)}^{3/2}}}C_{n}^{-1/2}(\mu )d\mu =-2\frac{d{{I}_{3}}}{d\xi } \\
=&-\frac{2\sqrt{2}}{n(n-1)}\left( \frac{\frac{{{(-1)}^{n}}}{2}{{e}^{{{(-1)}^{n}}\xi /2}}+(n-\frac{1}{2}){{e}^{-(n-1/2)\xi }}}{\sinh \xi }-\frac{\cosh \xi }{\sinh \xi }\frac{{{e}^{{{(-1)}^{n}}\xi /2}}-{{e}^{-(n-1/2)\xi }}}{\sinh \xi } \right) \\
{{I}_{5}}=&\int\limits_{-1}^{1}{1\cdot \frac{\cosh \xi -\mu }{1-{{\mu }^{2}}}}\frac{1}{{{\left( \cosh \xi -\mu  \right)}^{1/2}}}C_{n}^{-1/2}(\mu )d\mu =\frac{1}{2}\int{{{I}_{3}}\sinh \xi d\xi }\\
=&\frac{\sqrt{2}}{2n(n-1)}\left( \frac{{{e}^{{{(-1)}^{n}}\xi /2}}}{\frac{{{(-1)}^{n}}}{2}}+\frac{{{e}^{-(n-1/2)\xi }}}{(n-\frac{1}{2})} \right) \\
{{I}_{6}}=&\int\limits_{-1}^{1}{1\cdot \frac{\cosh \xi -\mu }{1-{{\mu }^{2}}}}\frac{\sinh \xi }{{{\left( \cosh \xi -\mu  \right)}^{3/2}}}C_{n}^{-1/2}(\mu )d\mu  \\
=&{{I}_{3}}\sinh \xi =\frac{\sqrt{2}}{n(n-1)}\left( {{e}^{{{(-1)}^{n}}\xi /2}}-{{e}^{-(n-1/2)\xi }} \right) \\
\end{aligned}
\end{equation}
If we integrate the terms for $n=0$ and $n=1$ , we get diverging integrals. Thus, instead of bracket expansion, we integrate the sum of two terms of the series
\begin{equation}\label{Intn0plusn1}
\int\limits_{\zeta =0}^{\pi }{\left. \frac{d}{d\xi }\left( \frac{{{U}_{0}}(\xi )C_{0}^{-1/2}(\mu )+{{U}_{1}}(\xi )C_{1}^{-1/2}(\mu )}{{{\left( \cosh \xi -\cos \zeta  \right)}^{1/2}}} \right)\psi \frac{1}{\rho } \right|_{\xi =-{{\tau }_{2}}}^{\xi ={{\tau }_{1}}}d\zeta }.
\end{equation}

\section{Transformations of kinetic energy coefficients \label{AppendixA1E}}
For the coefficients $B,{{C}_{11}},{{C}_{12}}$ and ${{D}_{1}}$ after integration along the contour (\ref{KineticIntegrContour}) we obtain formulas as in (\ref{KinEn}). But for coefficients ${{A}_{1}}$ and $E$ the following equalities were used.

After integration (\ref{KineticIntegrContour}) ${{A}_{1}}$ has the following form
\begin{equation}\label{A1before}
\begin{aligned}
   {{A}_{1}}=&\frac{1}{3}{{c}^{3}}\sum\limits_{n=2}^{\infty }{n(n-1){{e}^{-(2n-1){{\tau }_{1}}}}\left( 3\coth (n-1/2)({{\tau }_{1}}+{{\tau }_{2}})-1 \right)} \\
  =&{{c}^{3}}\sum\limits_{n=2}^{\infty }{n(n-1){{e}^{-(2n-1){{\tau }_{1}}}}\left( \frac{2}{3}+\frac{2}{{{e}^{(2n-1)({{\tau }_{1}}+{{\tau }_{2}})}}-1} \right)} \\
\end{aligned}
\end{equation}
Considering that \cite{bentwich1978exact}
\begin{equation}\label{sinht1powerminus3}
\sum\limits_{n=2}^{\infty }{n(n-1){{e}^{-(2n-1){{\tau }_{1}}}}}=\frac{1}{4{{\sinh }^{3}}{{\tau }_{1}}},
\end{equation}
we obtained for ${{A}_{1}}$ the following equality
\begin{equation}\label{A1after}
{{A}_{1}}=\frac{R_{1}^{3}}{6}+{{c}^{3}}\sum\limits_{n=2}^{\infty }{\frac{{{e}^{-\left( 2n-1 \right){{\tau }_{1}}}}}{{{e}^{\left( 2n-1 \right)\left( {{\tau }_{1}}+{{\tau }_{2}} \right)}}-1}}\frac{{{(2n-1)}^{2}}-1}{2}
\end{equation}
 	      	
After integration (\ref{KineticIntegrContour}) $E$ has the following form
\begin{equation}\label{Ebefore}
\begin{aligned}
   E=&{{c}^{3}}{{e}^{-({{\tau }_{1}}+{{\tau }_{2}})}}\frac{2+\coth{{\tau }_{1}}+\coth {{\tau }_{2}}}{2\sinh ({{\tau }_{1}})\sinh({{\tau }_{2}})}+{{c}^{3}}\sum\limits_{n=2}^{\infty }{\frac{{{e}^{-2n({{\tau }_{1}}+{{\tau }_{2}})}}}{({{e}^{(2n-1)(\tau_1+\tau_2)}}-1)4n(n-1){{\sinh }^{2}}{{\tau }_{1}}{{\sinh }^{2}}{{\tau }_{2}}} } \\
 & \times \Big(  \left( {{e}^{2n{{\tau }_{1}}}}-n{{e}^{2{{\tau }_{1}}}}+n-1 \right)\left( {{e}^{2n{{\tau }_{2}}}}+(-n{{e}^{2{{\tau }_{2}}}}+n-1){{e}^{(2n-1)({{\tau }_{1}}+{{\tau }_{2}})}} \right)  \\
 & + \left( {{e}^{2n{{\tau }_{2}}}}-n{{e}^{2{{\tau }_{2}}}}+n-1 \right)\left( {{e}^{2n{{\tau }_{1}}}}+(-n{{e}^{2{{\tau }_{1}}}}+n-1){{e}^{(2n-1)({{\tau }_{1}}+{{\tau }_{2}})}} \right) \Big) \\
\end{aligned}
\end{equation}
Taking into account the forms of ${{A}_{1}},B,{{C}_{11}},{{C}_{12}}$ and ${{D}_{1}}$, and symmetry considerations, we study the difference
\begin{equation}\label{Eminus}
E-{{c}^{3}}\sum\limits_{n=2}^{\infty }{\frac{{{e}^{-(2n-1)({{\tau }_{1}}+{{\tau }_{2}})}}}{{{e}^{\left( 2n-1 \right)\left( {{\tau }_{1}}+{{\tau }_{2}} \right)}}-1}}\frac{2\,{{S}_{n}}({{\tau }_{1}}){{S}_{n}}({{\tau }_{2}})}{{{(2n-1)}^{2}}-1}.
\end{equation}
This infinite sum may be calculated as follows
\begin{equation}\label{Eafter}
\begin{aligned}
  & E-{{c}^{3}}\sum\limits_{n=2}^{\infty }{\frac{{{e}^{-(2n-1)({{\tau }_{1}}+{{\tau }_{2}})}}}{{{e}^{\left( 2n-1 \right)\left( {{\tau }_{1}}+{{\tau }_{2}} \right)}}-1}}\frac{2\,{{S}_{n}}({{\tau }_{1}}){{S}_{n}}({{\tau }_{2}})}{{{(2n-1)}^{2}}-1}={{c}^{3}}{{e}^{-({{\tau }_{1}}+{{\tau }_{2}})}}\frac{2+\coth{{\tau }_{1}}+\coth {{\tau }_{2}}}{2\sinh ({{\tau }_{1}})\sinh({{\tau }_{2}})} \\
 & -{{c}^{3}}\frac{\coth ({{\tau }_{1}}+{{\tau }_{2}})-1}{4}\left( \frac{1}{{{\sinh }^{2}}{{\tau }_{1}}}+\frac{1}{{{\sinh }^{2}}{{\tau }_{2}}} \right)=\frac{{{c}^{3}}}{\sinh {{\tau }_{1}}\sinh {{\tau }_{2}}\sinh ({{\tau }_{1}}+{{\tau }_{2}})}=\frac{{{({{R}_{1}}{{R}_{2}})}^{2}}}{r}, \\
\end{aligned}
\end{equation}
and we obtain the same form as in (\ref{KinEn}).

\section{Coefficients' identity \label{AppendixEqualCoef}}
According to \cite{hicks1880,Voinov1969PMM,Voinov1970,petrov2011forced} , the kinetic energy can be presented as follows
\begin{equation}\label{AppendixKinEnHicksVoinov}
\begin{aligned}
   T=&2\pi {{\rho }_{l}}\big( {{A}_{1}}u_{1}^{2}+2B{{u}_{1}}{{u}_{2}}+{{A}_{2}}u_{2}^{2}+{{D}_{1}}{{{\dot{R}}}_{1}}^{2}+2E{{{\dot{R}}}_{1}}{{{\dot{R}}}_{2}}+{{D}_{2}}{{{\dot{R}}}_{2}}^{2}\\
 & +{{C}_{11}}{{u}_{1}}{{{\dot{R}}}_{1}}+{{C}_{12}}{{u}_{1}}{{{\dot{R}}}_{2}}+{{C}_{21}}{{u}_{2}}{{{\dot{R}}}_{1}}+{{C}_{22}}{{u}_{2}}{{{\dot{R}}}_{2}} \big), \\
  {{A}_{1}}=&\frac{R_{1}^{3}}{6}+\frac{1}{2}\sum\limits_{j=1}^{\infty }{{{\left( \frac{{{R}_{1}}}{A_{j}^{1}} \right)}^{3}}}, \qquad B=\frac{1}{2}\sum\limits_{j=1}^{\infty }{{{\left( \frac{{{R}_{2}}}{B_{j}^{1}} \right)}^{3}}}. \\
 {{C}_{11}}=&\sum\limits_{j=1}^{\infty }{\frac{R_{1}^{3}}{{{\left( A_{j}^{1} \right)}^{2}}B_{j}^{2}}}, \qquad \qquad {{C}_{12}}=\sum\limits_{j=1}^{\infty }{\frac{R_{1}^{3}}{{{\left( B_{j}^{2} \right)}^{2}}A_{j-1}^{1}}},\\
 {{D}_{1}}=&R_{1}^{3}+\sum\limits_{j=1}^{\infty }{\frac{R_{1}^{3}}{A_{j}^{1}}\left[ 1+({{(B_{j}^{2})}^{2}}-1)\ln \left( 1-\frac{1}{{{(B_{j}^{2})}^{2}}} \right) \right]},\\
E=&\frac{{{({{R}_{1}}{{R}_{2}})}^{2}}}{r}+\frac{{{({{R}_{1}}{{R}_{2}})}^{2}}}{{{R}_{2}}}\sum\limits_{j=1}^{\infty }{\left( \frac{1}{B_{j+1}^{2}}-B_{j}^{1}\ln \left( 1+\frac{1}{B_{j+1}^{2}B_{j}^{1}} \right) \right)},\\
\end{aligned}
\end{equation}
where $A_{j}^{i},B_{j}^{i}$ can be found using recurrent equations \cite{Voinov1969PMM}
\begin{equation}\label{AppendixrecurrVoinov}
\begin{aligned}
B_{j}^{i}=&\frac{r}{{{R}_{i}}}A_{j-1}^{k}-\frac{{{R}_{k}}}{{{R}_{i}}}B_{j-1}^{i},\\
A_{j}^{i}=&\frac{r}{{{R}_{i}}}B_{j}^{k}-\frac{{{R}_{k}}}{{{R}_{i}}}A_{j-1}^{i},
\end{aligned}
\end{equation}
with initial conditions $A_{0}^{i}=1, B_{0}^{i}=0$. These recurrent equations can be solved as follows \cite{Voinov1969PMM}
\begin{equation}\label{AjBjbyVoinov}
A_{j}^{i}=\frac{{{f}^{j+1}}-{{f}^{-(j+1)}}+\left( {{f}^{j}}-{{f}^{-j}} \right){{R}_{i}}/{{R}_{k}}}{f-{{f}^{-1}}},\enspace B_{j}^{i}=\frac{{{f}^{j}}-{{f}^{-j}}}{f-{{f}^{-1}}}\frac{r}{{{R}_{i}}}, \enspace i\ne k,  \enspace i,k=1,2,
\end{equation}
where $f$ is the root of
\begin{equation}\label{froot}
{{r}^{2}}f=\left( {{R}_{1}}f+{{R}_{2}} \right)\left( {{R}_{1}}+{{R}_{2}}f \right).
\end{equation}

It turns out that $f$   can be expressed through  ${{\tau }_{1}}$ and ${{\tau }_{2}}$ as $f={{e}^{-({{\tau }_{1}}+{{\tau }_{2}})}}$. Denote ${{q}_{i}}={{e}^{-{{\tau }_{i}}}}$. Then for $A_{j}^{i},B_{j}^{i}$ we obtain the following equalities
\[A_{j}^{i}=\frac{q_{i}^{-1}{{f}^{-j}}-{{q}_{i}}{{f}^{j}}}{q_{i}^{-1}-{{q}_{i}}}=\frac{\sinh ({{\tau }_{i}}+j({{\tau }_{1}}+{{\tau }_{2}}))}{\sinh {{\tau }_{i}}},\quad B_{j}^{i}=\frac{{{f}^{-j}}-{{f}^{j}}}{q_{k}^{-1}-{{q}_{k}}}=\frac{\sinh j({{\tau }_{1}}+{{\tau }_{2}})}{\sinh {{\tau }_{k}}}.\]
Thus, we obtained the same form for  ${{A}_{1}},{{A}_{2}},B$, as the one presented by Hicks \cite{hicks1880}.

According to the algorithm from \cite{neumann1883hydrodynamische}, we transform the kinetic energy coefficients as follows (taking into account that  $c={{R}_{1}}\sinh {{\tau }_{1}}={{R}_{2}}\sinh {{\tau }_{2}}$):
\begin{equation}\label{A1equal}
\begin{aligned}
{{A}_{1}}=&\frac{R_{1}^{3}}{6}+\frac{1}{2}\sum\limits_{j=1}^{\infty }{{{\left( \frac{c}{\sinh ({{\tau }_{1}}+j({{\tau }_{1}}+{{\tau }_{2}}))} \right)}^{3}}}=\frac{R_{1}^{3}}{6}+4{{c}^{3}}\sum\limits_{j=1}^{\infty }{{{\left( \frac{1}{q_{1}^{-1}{{f}^{-j}}-{{q}_{1}}{{f}^{j}}} \right)}^{3}}} \\
 =&\frac{R_{1}^{3}}{6}+4{{c}^{3}}\sum\limits_{j=1}^{\infty }{{{\left( \frac{{{q}_{1}}{{f}^{j}}}{1-q_{1}^{2}{{f}^{2j}}} \right)}^{3}}}=\frac{R_{1}^{3}}{6}+4{{c}^{3}}\sum\limits_{j=1}^{\infty }{\sum\limits_{n=0}^{\infty }{\frac{(n+1)(n+2)}{2}{{\left( {{q}_{1}}{{f}^{j}} \right)}^{2n+3}}}} \\
 =&\frac{R_{1}^{3}}{6}+4{{c}^{3}}\sum\limits_{j=1}^{\infty }{\sum\limits_{n=2}^{\infty }{\frac{n(n-1)}{2}{{\left( {{q}_{1}}{{f}^{j}} \right)}^{2n-1}}}}=\frac{R_{1}^{3}}{6}+4{{c}^{3}}\sum\limits_{n=2}^{\infty }{\frac{n(n-1)}{2}\frac{{{\left( {{q}_{1}}f \right)}^{2n-1}}}{1-{{f}^{2n-1}}}} \\
 =&\frac{R_{1}^{3}}{6}+4{{c}^{3}}\sum\limits_{n=2}^{\infty }{\frac{n(n-1)}{2}\frac{q_{1}^{2n-1}}{{{f}^{-(2n-1)}}-1}}=\frac{R_{1}^{3}}{6}+{{c}^{3}}\sum\limits_{n=2}^{\infty }{\frac{{{e}^{-\left( 2n-1 \right){{\tau }_{1}}}}}{{{e}^{\left( 2n-1 \right)\left( {{\tau }_{1}}+{{\tau }_{2}} \right)}}-1}}\frac{{{(2n-1)}^{2}}-1}{2}. \\
\end{aligned}
\end{equation}
Analogically, for $B$ we get that
\begin{equation}\label{Bequal}
B=\frac{1}{2}\sum\limits_{j=1}^{\infty }{{{\left( \frac{c}{\sinh j({{\tau }_{1}}+{{\tau }_{2}})} \right)}^{3}}}={{c}^{3}}\sum\limits_{n=2}^{\infty }{\frac{1}{{{e}^{\left( 2n-1 \right)\left( {{\tau }_{1}}+{{\tau }_{2}} \right)}}-1}\frac{{{(2n-1)}^{2}}-1}{2}}.
\end{equation}
Coefficient ${{C}_{11}}$ can be transformed as follows
\begin{equation}\label{C11equal}
\begin{aligned}
   {{C}_{11}}=&\sum\limits_{j=1}^{\infty }{\frac{R_{1}^{3}}{{{\left( A_{j}^{1} \right)}^{2}}B_{j}^{2}}}=\sum\limits_{j=1}^{\infty }{\frac{{{c}^{3}}}{{{\left( \sinh ({{\tau }_{1}}+j({{\tau }_{1}}+{{\tau }_{2}})) \right)}^{2}}\sinh j({{\tau }_{1}}+{{\tau }_{2}})}} \\
  =&8{{c}^{3}}\sum\limits_{j=1}^{\infty }{\frac{1}{\left( {{f}^{-j}}-{{f}^{j}} \right){{\left( q_{1}^{-1}{{f}^{-j}}-{{q}_{1}}{{f}^{j}} \right)}^{2}}}} \\
  =&8{{c}^{3}}\sum\limits_{j=1}^{\infty }{q_{1}^{2}{{f}^{3j}}\sum\limits_{n=0}^{\infty }{\frac{\left( 1-(n+2)q_{1}^{2(n+1)}+(n+1)q_{1}^{2(n+2)} \right)}{{{\left( 1-q_{1}^{2} \right)}^{2}}}{{\left( {{f}^{2n}} \right)}^{j}}}} \\
  =&8{{c}^{3}}\sum\limits_{n=0}^{\infty }{\frac{q_{1}^{-2}\left( 1-(n+2)q_{1}^{2(n+1)}+(n+1)q_{1}^{2(n+2)} \right)}{{{\left( q_{1}^{-2}-1 \right)}^{2}}}\frac{{{f}^{2n+3}}}{1-{{f}^{2n+3}}}} \\
  =&8{{c}^{3}}\sum\limits_{n=2}^{\infty }{\frac{q_{1}^{2n-2}\left( q_{1}^{-2n}-nq_{1}^{-2}+(n-1) \right)}{{{\left( q_{1}^{-2}-1 \right)}^{2}}}\frac{{{f}^{2n-1}}}{1-{{f}^{2n-1}}}}=8{{c}^{3}}\sum\limits_{n=2}^{\infty }{\frac{{{S}_{n}}({{e}^{2{{\tau }_{1}}}})\,{{e}^{-(2n-2){{\tau }_{1}}}}}{{{e}^{\left( 2n-1 \right)\left( {{\tau }_{1}}+{{\tau }_{2}} \right)}}-1}}. \\
\end{aligned}
\end{equation}

Analogically, for  ${{C}_{12}}$ we obtain that
\begin{equation}\label{C12equal}
{{C}_{12}}=\sum\limits_{j=1}^{\infty }{\frac{R_{1}^{3}}{{{\left( B_{j}^{2} \right)}^{2}}A_{j-1}^{1}}}=8{{c}^{3}}\sum\limits_{n=2}^{\infty }{\frac{{{S}_{n}}({{e}^{2{{\tau }_{2}}}})\,{{e}^{{{\tau }_{2}}}}}{{{e}^{\left( 2n-1 \right)\left( {{\tau }_{1}}+{{\tau }_{2}} \right)}}-1}}.
\end{equation}
The coefficient  ${{D}_{1}}$  can be transformed as follows
\begin{equation}\label{D1equal}
\begin{aligned}
   {{D}_{1}}=&R_{1}^{3}+\sum\limits_{j=1}^{\infty }{\frac{R_{1}^{3}}{A_{j}^{1}}\left[ 1+({{(B_{j}^{2})}^{2}}-1)\ln \left( 1-\frac{1}{{{(B_{j}^{2})}^{2}}} \right) \right]} \\
  =&R_{1}^{3}+8{{c}^{3}}\sum\limits_{j=1}^{\infty }{\frac{1}{{{(q_{1}^{-1}-{{q}_{1}})}^{2}}(q_{1}^{-1}{{f}^{-j}}-{{q}_{1}}{{f}^{j}})} } \\
 & \times \left[ 1+\left( {{\left( \frac{{{f}^{-j}}-{{f}^{j}}}{q_{1}^{-1}-{{q}_{1}}} \right)}^{2}}-1 \right)\ln \left( 1-{{\left( \frac{q_{1}^{-1}-{{q}_{1}}}{{{f}^{-j}}-{{f}^{j}}} \right)}^{2}} \right) \right] \\
 =&R_{1}^{3}+8{{c}^{3}}\sum\limits_{j=1}^{\infty }{\sum\limits_{n=1}^{\infty }{\frac{q_{1}^{-2n+3}{{\left( 1-(n+1)q_{1}^{2n}+nq_{1}^{2(n+1)} \right)}^{2}}}{n(n+1){{(1-q_{1}^{2})}^{4}}}}}{{\left( {{f}^{j}} \right)}^{2n+1}} \\
  =&R_{1}^{3}+8{{c}^{3}}\sum\limits_{n=1}^{\infty }{\frac{q_{1}^{-2n+3}q_{1}^{4(n+1)}{{\left( q_{1}^{-2(n+1)}-(n+1)q_{1}^{2}+n \right)}^{2}}}{n(n+1){{(1-q_{1}^{2})}^{4}}}}\frac{{{f}^{2n+1}}}{1-{{f}^{2n+1}}}\\
  =&R_{1}^{3}+8{{c}^{3}}\sum\limits_{n=2}^{\infty }{\frac{q_{1}^{-2n+5}{{\left( q_{1}^{-2n}-nq_{1}^{2}+n-1 \right)}^{2}}}{n(n-1){{(1-q_{1}^{2})}^{4}}}}\frac{1}{{{f}^{-(2n-1)}}-1}\\
  =&R_{1}^{3}+{{c}^{3}}\sum\limits_{n=2}^{\infty }{\frac{{{e}^{-(2n-1){{\tau }_{1}}}}}{{{e}^{\left( 2n-1 \right)\left( {{\tau }_{1}}+{{\tau }_{2}} \right)}}-1}}\frac{2\,S_{n}^{2}({{\tau }_{1}})}{{{(2n-1)}^{2}}-1}. \\
\end{aligned}
\end{equation}
Analogically, for $E$ we get
\begin{equation}\label{Eequal}
\begin{aligned}
E=&\frac{{{({{R}_{1}}{{R}_{2}})}^{2}}}{r}+\frac{{{({{R}_{1}}{{R}_{2}})}^{2}}}{{{R}_{2}}}\sum\limits_{j=1}^{\infty }{\left( \frac{1}{B_{j+1}^{2}}-B_{j}^{1}\ln \left( 1+\frac{1}{B_{j+1}^{2}B_{j}^{1}} \right) \right)} \\
=&\frac{{{({{R}_{1}}{{R}_{2}})}^{2}}}{r}+{{c}^{3}}\sum\limits_{n=2}^{\infty }{\frac{{{e}^{-(2n-1)({{\tau }_{1}}+{{\tau }_{2}})}}}{{{e}^{\left( 2n-1 \right)\left( {{\tau }_{1}}+{{\tau }_{2}} \right)}}-1}}\frac{2\,{{S}_{n}}({{\tau }_{1}}){{S}_{n}}({{\tau }_{2}})}{{{(2n-1)}^{2}}-1}. \\
\end{aligned}
\end{equation}
Thus, we prove that series  (\ref{KinEn}) coincide with Hicks's series  (\ref{KinEnHicks}) and Voinov's series (\ref{KinEnVoinov}).

\section{Expansion of kinetic energy coefficients by inverse powers $r$ \label{AppendixKinEnSeries}}
To compare the kinetic energy coefficients (\ref{KinEn}) with the ones in \cite{doinikov2015theoretical} series  (\ref{KinEn}) are considered in the inverse powers of  $r$ up to $o({{r}^{-10}})$:
\begin{equation}\label{KinEnSeries}
\begin{aligned}
 {{A}_{1}}=&\frac{R_{1}^{3}}{6}+\frac{R_{1}^{6}R_{2}^{3}}{2{{r}^{6}}}+\frac{3R_{1}^{6}R_{2}^{5}}{2{{r}^{8}}}+\frac{3R_{1}^{6}R_{2}^{7}}{{{r}^{10}}}, \\
 B=&\frac{R_{1}^{3}R_{2}^{3}}{2{{r}^{3}}}+\frac{R_{1}^{6}R_{2}^{6}}{2{{r}^{9}}}, \\
 {{A}_{2}}=&\frac{R_{2}^{3}}{6}+\frac{R_{1}^{3}R_{2}^{6}}{2{{r}^{6}}}+\frac{3R_{1}^{5}R_{2}^{6}}{2{{r}^{8}}}+\frac{3R_{1}^{7}R_{2}^{6}}{{{r}^{10}}}, \\
 {{C}_{11}}=&\frac{R_{1}^{5}R_{2}^{3}}{{{r}^{5}}}+\frac{2R_{1}^{5}R_{2}^{5}}{{{r}^{7}}}+\frac{3R_{1}^{5}R_{2}^{7}}{{{r}^{9}}}, \\
 {{C}_{12}}=&\frac{R_{1}^{3}R_{2}^{2}}{{{r}^{2}}}+\frac{R_{1}^{6}R_{2}^{5}}{{{r}^{8}}}+\frac{2R_{1}^{8}R_{2}^{5}+3R_{1}^{6}R_{2}^{7}}{{{r}^{10}}}, \\
 {{C}_{21}}=&\frac{R_{1}^{2}R_{2}^{3}}{{{r}^{2}}}+\frac{R_{1}^{5}R_{2}^{6}}{{{r}^{8}}}+\frac{2R_{1}^{5}R_{2}^{8}+3R_{1}^{7}R_{2}^{6}}{{{r}^{10}}}, \\
 {{C}_{22}}=&\frac{R_{1}^{3}R_{2}^{5}}{{{r}^{5}}}+\frac{2R_{1}^{5}R_{2}^{5}}{{{r}^{7}}}+\frac{3R_{1}^{7}R_{2}^{5}}{{{r}^{9}}}, \\
 {{D}_{1}}=&R_{1}^{3}+\frac{R_{1}^{4}R_{2}^{3}}{2{{r}^{4}}}+\frac{2R_{1}^{4}R_{2}^{5}}{3{{r}^{6}}}+\frac{3R_{1}^{4}R_{2}^{7}}{4{{r}^{8}}}+\frac{\frac{1}{2}R_{1}^{7}R_{2}^{6}+\frac{4}{5}R_{1}^{4}R_{2}^{9}}{{{r}^{10}}}, \\
 E=&\frac{R_{1}^{2}R_{2}^{2}}{r}+\frac{R_{1}^{5}R_{2}^{5}}{2{{r}^{7}}}+\frac{R_{1}^{7}R_{2}^{5}+R_{1}^{5}R_{2}^{7}}{{{r}^{9}}}, \\
 {{D}_{2}}=&R_{2}^{3}+\frac{R_{1}^{3}R_{2}^{4}}{2{{r}^{4}}}+\frac{2R_{1}^{5}R_{2}^{4}}{3{{r}^{6}}}+\frac{3R_{1}^{7}R_{2}^{4}}{4{{r}^{8}}}+\frac{\frac{1}{2}R_{1}^{6}R_{2}^{7}+\frac{4}{5}R_{1}^{9}R_{2}^{4}}{{{r}^{10}}}.
\end{aligned}
\end{equation}
These expansions coincide with the corresponding expansions from \cite{selby1890}.

The kinetic energy in \cite{doinikov2015theoretical} is presented in the following form
\begin{equation}\label{KinEnDoinikov}
\begin{aligned}
   T=&2\pi {{\rho }_{l}}\left( R_{1}^{3}\left( \dot{R}_{1}^{2}+\frac{\dot{z}_{1}^{2}}{6} \right)+R_{1}^{3}\left( {{f}_{1}}\dot{R}_{1}^{2}+{{f}_{4}}{{{\dot{R}}}_{1}}{{{\dot{z}}}_{1}}+{{f}_{5}}\dot{z}_{1}^{2} \right) \right.\\
  & +R_{1}^{2}{{R}_{2}}\left( {{f}_{2}}{{{\dot{R}}}_{1}}{{{\dot{R}}}_{2}}+{{f}_{6}}{{{\dot{R}}}_{2}}{{{\dot{z}}}_{1}}+{{f}_{3}}{{{\dot{R}}}_{1}}{{{\dot{z}}}_{2}}-{{f}_{7}}{{{\dot{z}}}_{1}}{{{\dot{z}}}_{2}} \right)  \\
 &  +R_{2}^{3}\left( \dot{R}_{2}^{2}+\frac{\dot{z}_{2}^{2}}{6} \right)+R_{2}^{3}\left( {{g}_{1}}\dot{R}_{2}^{2}-{{g}_{4}}{{{\dot{R}}}_{2}}{{{\dot{z}}}_{2}}+{{g}_{5}}\dot{z}_{2}^{2} \right)\\
 &\left. +{{R}_{1}}R_{2}^{2}\left( {{g}_{2}}{{{\dot{R}}}_{1}}{{{\dot{R}}}_{2}}+{{g}_{3}}{{{\dot{R}}}_{2}}{{{\dot{z}}}_{1}}+{{g}_{6}}{{{\dot{R}}}_{1}}{{{\dot{z}}}_{2}}+{{g}_{7}}{{{\dot{z}}}_{1}}{{{\dot{z}}}_{2}} \right) \right). \\
\end{aligned}
\end{equation}
Here we preserve the notations from \cite{doinikov2015theoretical}. Substituting the expressions for ${{f}_{i}}$, ${{g}_{i}}$ from Supplemental Material of \cite{doinikov2015theoretical}, and substituting ${{\dot{z}}_{1}}\to {{u}_{1}},\,{{\dot{z}}_{2}}\to -{{u}_{2}},\,\,D\to r$, we obtain the expansion up to $o({{r}^{-10}})$ for the kinetic energy coefficients
\begin{equation}\label{KinEnSeriesDoinikov}
\begin{aligned}
  {{A}_{1}}=&\frac{R_{1}^{3}}{6}+\frac{R_{1}^{6}R_{2}^{3}}{2{{r}^{6}}}+\frac{R_{1}^{6}R_{2}^{5}}{{{r}^{8}}}+\frac{2R_{1}^{6}R_{2}^{7}}{{{r}^{10}}}, \\
 B=&\frac{R_{1}^{3}R_{2}^{3}}{2{{r}^{3}}}+\frac{R_{1}^{6}R_{2}^{6}}{2{{r}^{9}}}, \\
 {{A}_{2}}=&\frac{R_{2}^{3}}{6}+\frac{R_{1}^{3}R_{2}^{6}}{2{{r}^{6}}}+\frac{R_{1}^{5}R_{2}^{6}}{{{r}^{8}}}+\frac{2R_{1}^{7}R_{2}^{6}}{{{r}^{10}}}, \\
 {{C}_{11}}=&\frac{R_{1}^{5}R_{2}^{3}}{{{r}^{5}}}+\frac{5R_{1}^{5}R_{2}^{5}}{3{{r}^{7}}}+\frac{5R_{1}^{5}R_{2}^{7}}{2{{r}^{9}}}, \\
 {{C}_{12}}=&\frac{R_{1}^{3}R_{2}^{2}}{{{r}^{2}}}+\frac{R_{1}^{6}R_{2}^{5}}{{{r}^{8}}}+\frac{R_{1}^{6}R_{2}^{7}+R_{1}^{8}R_{2}^{5}}{{{r}^{10}}}, \\
 {{C}_{21}}=&\frac{R_{1}^{2}R_{2}^{3}}{{{r}^{2}}}+\frac{R_{1}^{5}R_{2}^{6}}{{{r}^{8}}}+\frac{R_{1}^{7}R_{2}^{6}+R_{1}^{5}R_{2}^{8}}{{{r}^{10}}}, \\
 {{C}_{22}}=&\frac{R_{1}^{3}R_{2}^{5}}{{{r}^{5}}}+\frac{5R_{1}^{5}R_{2}^{5}}{3{{r}^{7}}}+\frac{5R_{1}^{7}R_{2}^{5}}{2{{r}^{9}}}, \\
 {{D}_{1}}=&R_{1}^{3}+\frac{R_{1}^{4}R_{2}^{3}}{2{{r}^{4}}}+\frac{2R_{1}^{4}R_{2}^{5}}{3{{r}^{6}}}+\frac{3R_{1}^{4}R_{2}^{7}}{4{{r}^{8}}}+\frac{R_{1}^{7}R_{2}^{6}}{2{{r}^{10}}}, \\
 E=&\frac{R_{1}^{2}R_{2}^{2}}{r}+\frac{R_{1}^{5}R_{2}^{5}}{2{{r}^{7}}}+\frac{R_{1}^{7}R_{2}^{5}+R_{1}^{5}R_{2}^{7}}{2{{r}^{9}}}, \\
 {{D}_{2}}=&R_{2}^{3}+\frac{R_{1}^{3}R_{2}^{4}}{2{{r}^{4}}}+\frac{2R_{1}^{5}R_{2}^{4}}{3{{r}^{6}}}+\frac{3R_{1}^{7}R_{2}^{4}}{4{{r}^{8}}}+\frac{R_{1}^{6}R_{2}^{7}}{2{{r}^{10}}}. \\
\end{aligned}
\end{equation}
Note that the coefficients in \cite{doinikov2015theoretical} coincide with the exact coefficients up to ${{r}^{-6}}.$

\section{Asymptotic expansion of coefficients ${{D}_{1}},E$ \label{AppendixAsymptDE}}

To obtain asymptotic expansions at small separation distance, in (\ref{KinEn}) we open the brackets for the corresponding coefficients ${{D}_{1}},E$. Using the method described in \cite{raszillier1990optimal}, we obtain that
\begin{equation}\label{D1Mellin}
\begin{aligned}
   {{D}_{1}}=&R_{1}^{3}+\frac{2{{c}^{3}}}{{{\sinh }^{4}}({{\tau }_{1}})}\frac{1}{2\pi i}\int\limits_{\sigma -i\infty }^{\sigma +i\infty }{{{\left( {{\tau }_{1}}+{{\tau }_{2}} \right)}^{-s}}\Gamma (s)}{{Z}_{D1}}(s,{{\lambda }_{1}})ds+r_{{{D}_{1}}}^{m} \\
  =&R_{1}^{3}+\frac{2{{c}^{3}}}{{{\sinh }^{4}}({{\tau }_{1}})}\sum\limits_{k=-1}^{n}{\underset{s=-k}{\mathop{\text{res}}}\,}\left( {{({{\tau }_{1}}+{{\tau }_{2}})}^{-s}}\Gamma (s){{Z}_{D1}}(s,{{\lambda }_{1}}) \right)+r_{{{D}_{1}}}^{m}, \\
\end{aligned}
\end{equation}
\begin{equation}\label{EMellin}
\begin{aligned}
  E=&\frac{{{({{R}_{1}}{{R}_{2}})}^{2}}}{r}+\frac{2{{c}^{3}}}{{{\sinh }^{2}}({{\tau }_{1}}){{\sinh }^{2}}({{\tau }_{2}})}\frac{1}{2\pi i}\int\limits_{\sigma -i\infty }^{\sigma +i\infty }{{{\left( {{\tau }_{1}}+{{\tau }_{2}} \right)}^{-s}}\Gamma (s)}{{Z}_{E}}(s,{{\lambda }_{1}},{{\lambda }_{2}})ds+r_{E}^{m} \\
 =&\frac{{{({{R}_{1}}{{R}_{2}})}^{2}}}{r}+\frac{2{{c}^{3}}}{{{\sinh }^{2}}({{\tau }_{1}}){{\sinh }^{2}}({{\tau }_{2}})}\sum\limits_{k=-1}^{n}{\underset{s=-k}{\mathop{\text{res}}}\,}\left( {{({{\tau }_{1}}+{{\tau }_{2}})}^{-s}}\Gamma (s){{Z}_{E}}(s,{{\lambda }_{1}},{{\lambda }_{2}}) \right)+r_{E}^{m}, \\
\end{aligned}
\end{equation}
where
\begin{equation}\label{ZD1}
\begin{aligned}
   {{Z}_{D1}}(s,k,{{\lambda }_{1}})=&\zeta (s,1-{{\lambda }_{1}})H(s)-2\cosh ({{\tau }_{1}})\zeta (s)H(s)+{{\cosh }^{2}}({{\tau }_{1}})\zeta (s,1+{{\lambda }_{1}})H(s) \\
 & +\sinh (2{{\tau }_{1}})\zeta (s,1+{{\lambda }_{1}})H(s-1)-2\sinh ({{\tau }_{1}})\zeta (s)H(s-1)\\
 &+{{\sinh }^{2}}({{\tau }_{1}})\zeta (s,1+{{\lambda }_{1}})H(s-2), \\
 \end{aligned}
\end{equation}
\begin{equation}\label{ZE}
 \begin{aligned}
  {{Z}_{E}}(s,k,{{\lambda }_{1}},{{\lambda }_{2}})=&\zeta (s)H(s)-\cosh ({{\tau }_{1}})\zeta (s,1+{{\lambda }_{1}})H(s)-\cosh ({{\tau }_{2}})\zeta (s,1+{{\lambda }_{2}})H(s) \\
 & +\cosh ({{\tau }_{1}})\cosh ({{\tau }_{2}})\zeta (s,2)H(s)-\sinh ({{\tau }_{1}})\zeta (s,1+{{\lambda }_{1}})H(s-1)\\
 &-\sinh ({{\tau }_{2}})\zeta (s,1+{{\lambda }_{2}})H(s-1)+ \sinh ({{\tau }_{1}}+{{\tau }_{2}})\zeta (s,2)H(s-1)\\
 &+\sinh ({{\tau }_{1}})\sinh ({{\tau }_{2}})\zeta (s,2)H(s-2). \\
\end{aligned}
\end{equation}
Moreover, function
\begin{equation}\label{Hs}
H(s)=\sum\limits_{n=2}^{\infty }{{{\left( 2n-1 \right)}^{-s}}\frac{1}{{{\left( 2n-1 \right)}^{2}}-1}}
\end{equation}
differs fundamentally from $Z(s)=\sum\limits_{n=2}^{\infty }{{{\left( 2n-1 \right)}^{-s}}\left( {{(2n-1)}^{2}}-1 \right)}$, which appears in the asymptotic expansion of coefficients  ${{A}_{1}}$ and $B$, and thus the deduction of the asymptotic expansion of coefficients $D_1$ and $E$ is much more complex than the one for coefficients $A_1$ and $B$. Function $Z(s)$
can be presented as the difference of two zeta functions with the corresponding coefficients  (\ref{Zs}). This does not hold for function  $H(s)$, for which the following recurrent equations hold
\begin{equation}\label{Hsrecur}
\begin{aligned}
   H(s)=&\sum\limits_{n=2}^{\infty }{{{\left( 2n-1 \right)}^{-s}}\frac{1}{{{\left( 2n-1 \right)}^{2}}-1}}=\sum\limits_{n=2}^{\infty }{{{\left( 2n-1 \right)}^{-(s+2)}}\frac{{{\left( 2n-1 \right)}^{2}}-1+1}{{{\left( 2n-1 \right)}^{2}}-1}} \\
  =&\sum\limits_{n=2}^{\infty }{{{\left( 2n-1 \right)}^{-(s+2)}}\frac{1}{{{\left( 2n-1 \right)}^{2}}-1}+}\sum\limits_{n=2}^{\infty }{{{\left( 2n-1 \right)}^{-(s+2)}}}\\
 =&H(s+2)+\zeta (s+2)\left( 1-{{2}^{-(s+2)}} \right)-1. \\
\end{aligned}
\end{equation}

Taking into account this recurrent equation, and the facts that $H(0)=1/4$  and $H(1)=3/4-\ln 2$, we obtain that
\begin{equation}\label{D1Y}
{{D}_{1}}=R_{1}^{3}+\frac{2{{c}^{3}}}{{{\sinh }^{4}}({{\tau }_{1}})}\sum\limits_{k=-1}^{n}{\underset{s=-k}{\mathop{\text{res}}}\,}\left( \Gamma (s){{({{\tau }_{1}}+{{\tau }_{2}})}^{-s}}{{Y}_{D1}}(s,k,{{\lambda }_{1}}) \right)+r_{{{D}_{1}}}^{m},
\end{equation}
\begin{equation}\label{EY}
E=\frac{{{({{R}_{1}}{{R}_{2}})}^{2}}}{r}+\frac{2{{c}^{3}}}{{{\sinh }^{2}}({{\tau }_{1}}){{\sinh }^{2}}({{\tau }_{2}})}\sum\limits_{k=-1}^{n}{\underset{s=-k}{\mathop{\text{res}}}\,}\left( \Gamma (s){{({{\tau }_{1}}+{{\tau }_{2}})}^{-s}}{{Y}_{E}}(s,k,{{\lambda }_{1}},{{\lambda }_{2}}) \right)+r_{E}^{m},
\end{equation}
where
\begin{equation}\label{YD1}
\begin{aligned}
   {{Y}_{D1}}(s,k,{{\lambda }_{1}})=&\zeta (s,1-{{\lambda }_{1}})F(s,k)-2\cosh ({{\tau }_{1}})\zeta (s)F(s,k)+{{\cosh }^{2}}({{\tau }_{1}})\zeta (s,1+{{\lambda }_{1}})F(s,k) \\
 &+\sinh (2{{\tau }_{1}})\zeta (s,1+{{\lambda }_{1}})F(s-1,k+1)-2\sinh({{\tau }_{1}})\zeta (s)F(s-1,k+1)\\
 &+{{\sinh }^{2}}({{\tau }_{1}})\zeta (s,1+{{\lambda }_{1}})F(s-2,k+2), \\
 \end{aligned}
\end{equation}

 \begin{equation}\label{YE}
 \begin{aligned}
 {{Y}_{E}}(s,k,{{\lambda }_{1}},{{\lambda }_{2}})=&\zeta (s)F(s,k)-\cosh ({{\tau }_{1}})\zeta (s,1+{{\lambda }_{1}})F(s,k)-\cosh ({{\tau }_{2}})\zeta (s,1+{{\lambda }_{2}})F(s,k)\\
 & +\cosh ({{\tau }_{1}})\cosh({{\tau }_{2}})\zeta (s,2)F(s,k)-\sinh({{\tau }_{1}})\zeta (s,1+{{\lambda }_{1}})F(s-1,k+1)\\
 &-\sinh({{\tau }_{2}})\zeta (s,1+{{\lambda }_{2}})F(s-1,k+1)+ \sinh ({{\tau }_{1}}+{{\tau }_{2}})\zeta (s,2)F(s-1,k+1)\\
 &+\sinh ({{\tau }_{1}})\sinh({{\tau }_{2}})\zeta (s,2)F(s-2,k+2), \\
\end{aligned}
\end{equation}

\begin{equation}\label{Fs}
 F(s,k)=\left\{ \begin{aligned}
  & \frac{1}{4}-\frac{k}{2}+\sum\limits_{j=0}^{k/2-1}{\zeta (s+k-2j)(1-{{2}^{-(s+k-2j)}})},& k- \text{even}\\
 &\frac{1}{4}-\frac{k}{2}-\ln 2+\sum\limits_{j=1}^{(k-1)/2}{\zeta (s+k+1-2j)(1-{{2}^{-(s+k+1-2j)}})}& \\
 & +\zeta (s+k+1)(1-{{2}^{-(s+k+1)}}), & k- \text{odd}, \\
\end{aligned} \right.
\end{equation}

After calculating the residue, for ${{D}_{1}},E$ we get

  \begin{equation}\label{D1W}
  \begin{aligned}
  {{D}_{1}}=&R_{1}^{3}+\frac{2{{c}^{3}}}{{{\sinh }^{4}}({{\tau }_{1}})}\Bigg(  -\frac{\cosh {{\tau }_{1}}-1}{2({{\tau }_{1}}+{{\tau }_{2}})}\bigg( \psi (1+{{\lambda }_{1}})\\
   &  +\cosh ({{\tau }_{1}})(\psi (1+{{\lambda }_{1}})+\ln ({{\tau }_{1}}+{{\tau }_{2}})-1 +3\ln 2)+\ln \left( \frac{{{\tau }_{1}}+{{\tau }_{2}}}{2} \right)-\sinh {{\tau }_{1}}+2  \bigg)\\
  & +\sum\limits_{k=0}^{m}{{{(-1)}^{k}}\frac{{{({{\tau }_{1}}+{{\tau }_{2}})}^{k}}}{k!}{{W}_{D1}}(k,{{\lambda }_{1}})}  \Bigg) +r_{{{D}_{1}}}^{m}, \\
 \end{aligned}
  \end{equation}

 \begin{equation}\label{EW}
\begin{aligned}
E=&\frac{{{({{R}_{1}}{{R}_{2}})}^{2}}}{r}+\frac{2{{c}^{3}}}{{{\sinh }^{2}}({{\tau }_{1}}){{\sinh }^{2}}({{\tau }_{2}})}\Bigg( \frac{\sinh ({{\tau }_{1}})\sinh ({{\tau }_{2}})}{{{\tau }_{1}}+{{\tau }_{2}}} \left( -\frac{1}{2}\ln (2({{\tau }_{1}}+{{\tau }_{2}}))-\frac{3}{4}+\frac{\gamma }{2} \right)  \\
&+\frac{1}{4} \frac{\sinh ({{\tau }_{1}}+{{\tau }_{2}})-\sinh ({{\tau }_{1}})-\sinh ({{\tau }_{2}})}{\tau_1+\tau_2}+\left( \frac{3}{4}-\ln 2 \right)\frac{(\cosh ({{\tau }_{1}})-1)(\cosh ({{\tau }_{2}})-1)}{\tau_1+\tau_2} \\
 & +\left. \sum\limits_{k=0}^{m}{{{(-1)}^{k}}\frac{{{({{\tau }_{1}}+{{\tau }_{2}})}^{k}}}{k!}{{W}_{E}}(k,{{\lambda }_{1}})} \right)+r_{E}^{m} ,\\
 \end{aligned}
 \end{equation}
where
\begin{equation}\label{WD1}
\begin{aligned}
   {{W}_{D1}}(k,{{\lambda }_{1}})=&\zeta (-k,1-{{\lambda }_{1}})G(k)-2\cosh ({{\tau }_{1}})\zeta (-k)G(k)+{{\cosh }^{2}}({{\tau }_{1}})\zeta (-k,1+{{\lambda }_{1}})G(k) \\
 & +\sinh (2{{\tau }_{1}})\zeta (-k,1+{{\lambda }_{1}})G(k+1)-2\sinh ({{\tau }_{1}})\zeta (-k)G(k+1)\\
 &{{\sinh }^{2}}({{\tau }_{1}})\zeta (-k,1+{{\lambda }_{1}})G(k+2) \\
 & +\frac{1}{2}\left\{ \begin{aligned}
  &\begin{aligned} & \sinh (2{{\tau }_{1}})\left( {{\zeta }^{(1,0)}}(-k,1+{{\lambda }_{1}})+L(k)\zeta (-k,1+{{\lambda }_{1}}) \right)\\
  &-2\sinh ({{\tau }_{1}})\left( {{\zeta }^{(1,0)}}(-k,1)+L(k)\zeta (-k,1) \right),\end{aligned} & \begin{aligned}
   &  \\
  & k- \text{even},\end{aligned} \\
 &\begin{aligned} & \left( {{\zeta }^{(1,0)}}(-k,1-{{\lambda }_{1}})+L(k)\zeta (-k,1-{{\lambda }_{1}}) \right)\\
 &-2\cosh ({{\tau }_{1}})\left( {{\zeta }^{(1,0)}}(-k,1)+L(k)\zeta (-k,1) \right) \\
 & +\cosh (2{{\tau }_{1}})\left( {{\zeta }^{(1,0)}}(-k,1+{{\lambda }_{1}})+L(k)\zeta (-k,1+{{\lambda }_{1}}) \right), \end{aligned} & \begin{aligned}
   &  \\
   &  \\
  & k- \text{odd},\end{aligned} \\
\end{aligned} \right. \\
 \end{aligned}
\end{equation}

\begin{equation}\label{WE}
 \begin{aligned}
 {{W}_{E}}(s,k,{{\lambda }_{1}},{{\lambda }_{2}})=&\zeta (s)G(k)-\cosh ({{\tau }_{1}})\zeta (s,1+{{\lambda }_{1}})G(k)-\cosh ({{\tau }_{2}})\zeta (s,1+{{\lambda }_{2}})G(k) \\
 & +\cosh ({{\tau }_{1}})\cosh ({{\tau }_{2}})\zeta (s,2)G(k)-\sinh ({{\tau }_{1}})\zeta (s,1+{{\lambda }_{1}})G(k+1)\\
 &-\sinh ({{\tau }_{2}})\zeta (s,1+{{\lambda }_{2}})G(k+1)+ \sinh ({{\tau }_{1}}+{{\tau }_{2}})\zeta (s,2)G(k+1)\\
 &+\sinh ({{\tau }_{1}})\sinh ({{\tau }_{2}})\zeta (s,2)G(k+2) \\
 & +\frac{1}{2}\left\{ \begin{aligned}
  & \sinh ({{\tau }_{1}}+{{\tau }_{2}})\left( {{\zeta }^{(1,0)}}(-k,2)+L(k)\zeta (-k,2) \right)&\\
  &-\sinh ({{\tau }_{1}})\left( {{\zeta }^{(1,0)}}(-k,1+{{\lambda }_{1}})+L(k)\zeta (-k,1+{{\lambda }_{1}}) \right)& \\
 & -\sinh ({{\tau }_{2}})\left( {{\zeta }^{(1,0)}}(-k,1+{{\lambda }_{2}})+L(k)\zeta (-k,1+{{\lambda }_{2}}) \right),& k - \text{even} \\
 & \left( {{\zeta }^{(1,0)}}(-k,1)+L(k)\zeta (-k,1) \right)&\\
 &-\cosh ({{\tau }_{1}})\left( {{\zeta }^{(1,0)}}(-k,1+{{\lambda }_{1}})+L(k)\zeta (-k,1+{{\lambda }_{1}}) \right)& \\
 & -\cosh ({{\tau }_{2}})\left( {{\zeta }^{(1,0)}}(-k,1+{{\lambda }_{2}})+L(k)\zeta (-k,1+{{\lambda }_{2}}) \right)&\\
 &+\cosh ({{\tau }_{1}}+{{\tau }_{2}})\left( {{\zeta }^{(1,0)}}(-k,2)+L(k)\zeta (-k,2) \right),& k-\text{odd} \\
\end{aligned} \right. \\
\end{aligned}
\end{equation}

\begin{equation}\label{Gk}
 G(k)=\left\{ \begin{aligned}
  & \frac{1}{4}-\frac{k}{2},&k-\text{even} \\
 & \frac{1}{4}-\frac{k}{2}-\ln 2+\sum\limits_{j=1}^{(k-1)/2}{\zeta (1-2j)(1-{{2}^{-(1-2j)}})},&k- \text{odd}\\
\end{aligned} \right. \\
\end{equation}

\begin{equation}\label{Lk}
 L(k)={{H}_{k}}-\ln \frac{{{\tau }_{1}}+{{\tau }_{2}}}{2}, \quad {{H}_{k}}=\sum\limits_{i=1}^{k}{\frac{1}{i}}\\
\end{equation}

\section{Comparison of coefficients' asymptotic expansions \label{AppendixEqualDAN}}
To compare the asymptotic expansion of the kinetic energy obtained in this paper with the three-terms expansion from \cite{sanduleanu2018trinomial},  we pass from  ${{\tau }_{1}}+{{\tau }_{2}}$ to $h=r-{{R}_{1}}-{{R}_{2}}$
\begin{equation}\label{A1forDAN}
\begin{aligned}
{{A}_{1}}=&\frac{R_{1}^{3}}{6}+\frac{{{p}^{3}}}{2}\zeta \left( 3,1+{{\alpha }_{2}} \right)+\frac{{{p}^{2}}}{4}h\ln \frac{h}{2p}\\
 & +h\frac{{{p}^{2}}}{2}\bigg( \psi \left( 1+{{\alpha }_{2}} \right)+\left( \alpha _{1}^{3}+\alpha _{2}^{3} \right)\zeta \left( 3,1+{{\alpha }_{2}} \right)\\
 &+{{\alpha }_{1}}{{\alpha }_{2}}\left( {{\alpha }_{2}}-{{\alpha }_{1}} \right)\zeta \left( 4,1+{{\alpha }_{2}} \right)+\frac{1}{6} \bigg) \\
 \end{aligned}
\end{equation}
\begin{equation}\label{BforDAN}
\begin{aligned}
&B=\frac{{{p}^{3}}}{2}\zeta (3)+\frac{{{p}^{2}}}{4}h\ln \frac{h}{2p}+h\frac{{{p}^{2}}}{2}\left( \left( \alpha _{1}^{3}+\alpha _{2}^{3} \right)\zeta (3)-\gamma +\frac{1}{6} \right) \\
 \end{aligned}
\end{equation}
\begin{equation}\label{C11forDAN}
\begin{aligned}
{{C}_{11}}=&pR_{1}^{2}\left( \gamma +\psi \left( 1+{{\alpha }_{2}} \right)-{{\alpha }_{2}}\zeta \left( 2,1+{{\alpha }_{2}} \right) \right) +\frac{{{p}^{2}}}{2}h\ln \frac{h}{2p}\\
 & +h \left( \frac{{{p}^{2}}}{6}+\frac{R_{1}^{2}}{3}\left( \gamma \left( \alpha _{1}^{3}+\alpha _{2}^{3} \right)+ \right.\left( \alpha _{1}^{2}-{{\alpha }_{1}}{{\alpha }_{2}}+4\alpha _{2}^{2} \right)\psi \left( 1+{{\alpha }_{2}} \right)\right. \\
 &  \left. -{{\alpha }_{2}}\left( \alpha _{1}^{3}+{{\alpha }_{1}}\alpha _{2}^{2}+2\alpha _{2}^{2} \right)\zeta \left( 2,1+{{\alpha }_{2}} \right)-2{{\alpha }_{1}}\alpha _{2}^{2}\left( {{\alpha }_{2}}-{{\alpha }_{1}} \right)\zeta \left( 3,1+{{\alpha }_{2}} \right) \right)\!\! \bigg) , \\
 \end{aligned}
\end{equation}
\begin{equation}\label{C12forDAN}
\begin{aligned}
{{C}_{12}}=&-pR_{2}^{2}\left( \gamma +\psi \left( {{\alpha }_{2}} \right)+{{\alpha }_{1}}\zeta \left( 2 \right) \right) +\frac{{{p}^{2}}}{2}h\ln \frac{h}{2p}\\
 & +h \left(\frac{{{p}^{2}}}{6}+\frac{R_{2}^{2}}{3}\left( -\gamma \left( 4\alpha _{1}^{3}+3\alpha _{1}^{2}{{\alpha }_{2}}+\alpha _{2}^{3} \right)- \right.\left( \alpha _{1}^{3}+\alpha _{2}^{3} \right)\psi \left( {{\alpha }_{2}} \right) \right. \\
& \left. -2\left( \alpha _{1}^{3}+\alpha _{2}^{3} \right)\zeta \left( 2 \right)+{{\alpha }_{1}}{{\alpha }_{2}}\left( {{\alpha }_{2}}-{{\alpha }_{1}} \right)\zeta \left( 2,{{\alpha }_{2}} \right) \right) \!\! \bigg) , \\
 \end{aligned}
\end{equation}
\begin{equation}\label{D1forDAN}
\begin{aligned}
  {{D}_{1}}=&R_{1}^{3}\left( 1-\frac{{{\zeta }^{(1,0)}}(-1,1-{{\alpha }_{2}})+{{\zeta }^{(1,0)}}(-1,1+{{\alpha }_{2}})-2{{\zeta }^{(1,0)}}(-1,1)}{{{\alpha }_{2}}} \right.\\
 &+  2\ln (\Gamma (1+{{\alpha }_{2}}))-{{\alpha }_{2}}(\psi (1+{{\alpha }_{2}})-1) \bigg) +\frac{{{p}^{2}}}{4}h\ln \frac{h}{2p}\\
 & +h\left(\frac{{{p}^{2}}}{12}+ \frac{R_{1}^{2}p({{R}_{1}}-{{R}_{2}})\left( {{\zeta }^{(1,1)}}(-1,1-{{\alpha }_{2}})-{{\zeta }^{(1,1)}}(-1,1+{{\alpha }_{2}}) \right)}{3R_{2}^{2}} \right. \\
 &-\frac{R_{1}^{2}\left( {{\zeta }^{(1,0)}}(-3,1-{{\alpha }_{2}})+{{\zeta }^{(1,0)}}(-3,1+{{\alpha }_{2}}) \right)}{3\alpha _{2}^{2}}  \\
 & +R_{1}^{2}\left( \alpha _{1}^{3}+\alpha _{2}^{3} \right)\frac{\left( {{\zeta }^{(1,0)}}(-1,1-{{\alpha }_{2}})+{{\zeta }^{(1,0)}}(-1,1+{{\alpha }_{2}}) \right)}{3\alpha _{2}^{2}}\\
 &-4R_{1}^{2}{{\zeta }^{(1,0)}}(-1,1+{{\alpha }_{2}})+2R_{1}^{2}\frac{{{\zeta }^{(1,0)}}(-2,1+{{\alpha }_{2}})}{{{\alpha }_{2}}} \\
 & +R_{1}^{2}\frac{2}{\alpha _{2}^{2}}\left( \frac{1}{3}{\zeta }'(-3)-{{\alpha }_{2}}{\zeta }'(-2) \right)+2R_{1}^{2}{{\alpha }_{2}}\ln (\Gamma (1+{{\alpha }_{2}}))\\
 &+R_{1}^{2}\frac{\alpha _{1}^{2}-{{\alpha }_{1}}{{\alpha }_{2}}-\alpha _{2}^{2}}{3}\psi (1+{{\alpha }_{2}}) +R_{1}^{2}\frac{{{\alpha }_{1}}{{\alpha }_{2}}\left( {{\alpha }_{2}}-{{\alpha }_{1}} \right)}{3}\zeta (2,1+{{\alpha }_{2}})\\
 &-\left. R_{1}^{2}\frac{2}{\alpha _{2}^{2}}\left( \frac{1}{3}-{{\alpha }_{2}} \right){\zeta }'(-1)-R_{1}^{2}{{\alpha }_{2}}\ln \text{ }2\pi +R_{1}^{2}\frac{11\alpha _{1}^{2}+4{{\alpha }_{1}}{{\alpha }_{2}}+52\alpha _{2}^{2}}{36} \right) \\
\end{aligned}
\end{equation}
\begin{equation}\label{EforDAN}
\begin{aligned}
E=&\frac{R_{1}^{2}R_{2}^{2}}{{{R}_{1}}+{{R}_{2}}}\left( 1+\gamma  \right)-{{R}_{1}}{{R}_{2}}\left( {{R}_{1}}\ln \Gamma (1+{{\alpha }_{1}})+{{R}_{2}}\ln \Gamma (1+{{\alpha }_{2}}) \right) \\
&+{{R}_{1}}{{R}_{2}}\left( {{R}_{1}}+{{R}_{2}} \right)\left( {{\zeta }^{(1,0)}}(-1,1+{{\alpha }_{1}})+{{\zeta }^{(1,0)}}(-1,1+{{\alpha }_{2}})-2{{\zeta }^{(1,0)}}(-1,1) \right)\\
&+\frac{{{p}^{2}}}{4}h\ln \frac{h}{2p} +h \Bigg( \frac{{{p}^{2}}}{12}+ \frac{1}{3}p({{R}_{1}}-{{R}_{2}})\left( {{\zeta }^{(1,1)}}(-1,1+{{\alpha }_{2}})-{{\zeta }^{(1,1)}}(-1,1+{{\alpha }_{1}}) \right) \\
 & +{{({{R}_{1}}+{{R}_{2}})}^{2}} \bigg( \frac{1}{3} \left({{\zeta }^{(1,0)}}(-3,1+{{\alpha }_{1}})+{{\zeta }^{(1,0)}}(-3,1+{{\alpha }_{2}}) \right)\\
 &- \frac{1}{3} \left( {{\zeta }^{(1,0)}}(-1,1+{{\alpha }_{1}})+{{\zeta }^{(1,0)}}(-1,1+{{\alpha }_{2}}) \right) \\
 & -{{\alpha }_{1}}\left( {{\zeta }^{(1,0)}}(-2,1+{{\alpha }_{1}})-{{\zeta }^{(1,0)}}(-1,1+{{\alpha }_{1}}) \right)\\
 &-{{\alpha }_{2}}\left( {{\zeta }^{(1,0)}}(-2,1+{{\alpha }_{2}})-{{\zeta }^{(1,0)}}(-1,1+{{\alpha }_{2}}) \right) \bigg) \\
 & -{{R}_{1}}{{R}_{2}}\left( \frac{1}{3}({{\alpha }_{1}}-{{\alpha }_{2}})\left({{\alpha }_{1}}\psi (1+{{\alpha }_{1}})-{{\alpha }_{2}}\psi (1+{{\alpha }_{2}})\right) \right.\\
 &\left.+\frac{1}{3}\gamma \left( \alpha _{1}^{3}+\alpha _{2}^{3} \right)-2{\zeta }'(-1)+\frac{5}{12}-\frac{1}{2}\ln (2\pi ) \right)\\
 &+\frac{1}{3}{{({{R}_{1}}+{{R}_{2}})}^{2}}\left( {\zeta }'(-1)-3{\zeta }'(-2)+2{\zeta }'(-3) \right)-\frac{23{{p}^{2}}}{36} \Bigg) , \\
\end{aligned}
\end{equation}
where ${{\alpha }_{i}}=R_i/(R_1+R_2)$, $p=R_1 R_2/(R_1+R_2)$.

The expressions obtained coincide with the ones in \cite{sanduleanu2018trinomial}
\begin{equation}\label{A1DAN}
\begin{aligned}
{{A}_{1}}=&R_{1}^{3}\left( \frac{1}{6}+\frac{1}{2}\sum\limits_{n=1}^{\infty }{{{\left( \frac{{{\alpha }_{2}}}{n+{{\alpha }_{2}}} \right)}^{3}}} \right)+{{p}^{2}}h\left( \frac{1}{4}\left( \ln \left( \frac{h}{2p} \right)-1 \right)+\frac{1}{3}-\frac{\gamma }{2}\right.\\
  &\left. +\frac{1}{2}\sum\limits_{n=1}^{\infty }{\left( \frac{1}{n}-\frac{n(n+1)(n-1+3{{\alpha }_{2}})}{{{(n+{{\alpha }_{2}})}^{4}}} \right)} \right), \\
\end{aligned}
\end{equation}
\begin{equation}\label{BDAN}
\begin{aligned}
 & B=\frac{1}{2}{{p}^{3}}\zeta (3)+{{p}^{2}}h\left( \frac{1}{4}\left( \ln \left( \frac{h}{2p} \right)-1 \right)+\frac{1}{3}-\frac{\gamma }{2}+\frac{1}{2}(1-3{{\alpha }_{1}}{{\alpha }_{2}})\zeta (3) \right),\  \\
 \end{aligned}
\end{equation}
\begin{equation}\label{C11DAN}
\begin{aligned}
 {{C}_{11}}=&R_{1}^{3}\sum\limits_{n=1}^{\infty }{\frac{\alpha _{2}^{3}}{n{{(n+{{\alpha }_{2}})}^{2}}}}+2{{p}^{2}}h\left( \frac{1}{4}\left( \ln \left( \frac{h}{2p} \right)-1 \right)+\frac{1}{3}-\frac{\gamma }{2} \right. \\
 & +\left. \sum\limits_{n=1}^{\infty }{\frac{2{{\alpha }_{2}}{{n}^{2}}+3(\alpha _{1}^{3}+\alpha _{2}^{3}+\alpha _{2}^{2})n+{{\alpha }_{2}}(\alpha _{1}^{3}+\alpha _{2}^{3}+3\alpha _{2}^{2})}{6n{{(n+{{\alpha }_{2}})}^{3}}} } \right),\  \\
  \end{aligned}
\end{equation}
\begin{equation}\label{C12DAN}
\begin{aligned}
  {{C}_{12}}=&R_{1}^{3}\sum\limits_{n=1}^{\infty }{\frac{\alpha _{2}^{3}}{{{n}^{2}}(n-1+{{\alpha }_{2}})}}+2{{p}^{2}}h\left( \frac{1}{4}\left( \ln \left( \frac{h}{2p} \right)-1 \right)+\frac{1}{3}-\frac{\gamma }{2} \right.\\
 &\left. +\sum\limits_{n=1}^{\infty }{\left[ \frac{-{{\alpha }_{1}}{{n}^{2}}+(3n-2{{\alpha }_{1}})(\alpha _{1}^{3}+\alpha _{2}^{3})}{6{{n}^{2}}{{(n-{{\alpha }_{1}})}^{2}}} \right]} \right), \\
 \end{aligned}
\end{equation}
\begin{equation}\label{D1DAN}
\begin{aligned}
 {{D}_{1}}=&R_{1}^{3}+R_{1}^{3}\sum\limits_{n=1}^{\infty }{\frac{{{\alpha }_{2}}}{n+{{\alpha }_{2}}}}\left[ 1+\left( \frac{{{n}^{2}}}{\alpha _{2}^{2}}-1 \right)\ln \left( 1-\frac{\alpha _{2}^{2}}{{{n}^{2}}} \right) \right]+{{p}^{2}}h\left( \frac{1}{4}\left( \ln \left( \frac{h}{2p} \right)-1 \right) \right. +\frac{1}{3}-\frac{\gamma }{2}\\
 &\left. +\sum\limits_{n=1}^{\infty }{\left[ \frac{1}{6n}+\frac{(n-1)(n+1-3{{\alpha }_{2}})}{3n\alpha _{2}^{2}}\left( 1+\frac{{{n}^{2}}}{\alpha _{2}^{2}}\ln \left( 1-\frac{\alpha _{2}^{2}}{{{n}^{2}}} \right) \right)+\frac{\alpha _{1}^{2}-(n+1){{\alpha }_{2}}}{3n{{(n+{{\alpha }_{2}})}^{2}}} \right]} \right), \\
 \end{aligned}
\end{equation}
\begin{equation}\label{EDAN}
\begin{aligned}
  E=&\frac{{{({{R}_{1}}{{R}_{2}})}^{2}}}{{{R}_{1}}+{{R}_{2}}}\left( 1+\sum\limits_{n=1}^{\infty }{\frac{1}{n+1}}\left( 1-\frac{n(n+1)}{{{\alpha }_{1}}{{\alpha }_{2}}}\ln \left( 1+\frac{{{\alpha }_{1}}{{\alpha }_{2}}}{n(n+1)} \right) \right) \right) \\
 &+{{p}^{2}}h\left( \frac{1}{4}\left( \ln \left( \frac{h}{2p} \right)-1 \right)+ \right.\frac{1}{3}-\frac{\gamma }{2}-\frac{1}{2} \\
 & \left. +\sum\limits_{n=1}^{\infty }{\left( -\frac{1}{6(n+1)}+\frac{{{n}^{2}}-1+3{{\alpha }_{1}}{{\alpha }_{2}}}{3(n+1){{\alpha }_{1}}{{\alpha }_{2}}}\left( 1-\frac{n(n+1)}{{{\alpha }_{1}}{{\alpha }_{2}}}\ln \left( 1+\frac{{{\alpha }_{1}}{{\alpha }_{2}}}{n(n+1)} \right) \right) \right)} \right). \\
\end{aligned}
\end{equation}
The equality of expansions can be verified numerically or analytically. The equalities for  ${{A}_{1}},B$ were shown in \cite{raszillier1990optimal}. ${{C}_{11}},{{C}_{12}}$ can be calculated. For ${{D}_{1}},E$ the situation is more complicated. To obtain the equality for ${{D}_{1}}$, we use the already mentioned algorithm  \cite{neumann1883hydrodynamische} to convert the sum
\begin{equation}\label{D1X0DAN}\sum\limits_{n=1}^{\infty }{\frac{{{\alpha }_{2}}}{n+{{\alpha }_{2}}}}\left( 1+\left( \frac{{{n}^{2}}}{\alpha _{2}^{2}}-1 \right)\ln \left( 1-\frac{\alpha _{2}^{2}}{{{n}^{2}}} \right) \right)=\sum\limits_{j=2}^{\infty }{\sum\limits_{n=1}^{\infty }{\frac{(j-1)}{j}\left( \frac{\alpha _{2}^{2j-1}}{{{n}^{2j-1}}}-\frac{\alpha _{2}^{2j}}{{{n}^{2j}}} \right)}},
\end{equation}

\begin{equation}\label{D1X1DAN}
\begin{aligned}
  & \sum\limits_{n=1}^{\infty }{\left( \frac{1}{6n}+\frac{(n-1)(n+1-3{{\alpha }_{2}})}{3n\alpha _{2}^{2}}\left( 1+\frac{{{n}^{2}}}{\alpha _{2}^{2}}\ln \left( 1-\frac{\alpha _{2}^{2}}{{{n}^{2}}} \right) \right)+\frac{\alpha _{1}^{2}-(n+1){{\alpha }_{2}}}{3n{{(n+{{\alpha }_{2}})}^{2}}} \right)}= \\
 & =\sum\limits_{j=1}^{\infty }{\sum\limits_{n=1}^{\infty }{\left( \frac{\alpha _{2}^{2j-1}}{(j+1){{n}^{2j}}}+\frac{(j+2)\alpha _{2}^{2j-2}-3(j+2)\alpha _{2}^{2j-1}-(j+1)\alpha _{2}^{2j}}{3(j+1)(j+2){{n}^{2j+1}}} \right)}} .\\
 \end{aligned}
\end{equation}
The double sums obtained must be calculated first by  $n$, and then by $j$.

\end{document}